\documentclass[twocolumn]{aastex63} 

\usepackage{amsmath}
\usepackage{amssymb}
\usepackage{empheq}
\usepackage{mathrsfs}
\usepackage{enumitem}
				
\usepackage{amssymb}

\shorttitle{An Outer Companion Orbiting 51 Peg?}
\shortauthors{Morgan et al.}
\graphicspath{{./}{figures/}}
\begin{document}


\title{An Outer Giant Planet or Brown Dwarf in the 51 Pegasi System?}

\correspondingauthor{Marvin Morgan}
\email{marvinmorgan@ucsb.edu}

\author[0000-0003-4022-6234]{Marvin Morgan}
\affiliation{Department of Physics, University of California, Santa Barbara, Santa Barbara, CA 93106, USA}

\author[0000-0003-2649-2288]{Brendan P. Bowler}
\affiliation{Department of Physics, University of California, Santa Barbara, Santa Barbara, CA 93106, USA}

\author[0000-0003-4557-414X]{Kyle Franson}
\altaffiliation{NHFP Sagan Fellow}
\affiliation{Department of Astronomy $\&$ Astrophysics, University of California, Santa Cruz, CA 95064, USA}

\author[0000-0003-4006-102X]{Lillian Jiang}
\affiliation{Department of Physics, University of California, Santa Barbara, Santa Barbara, CA 93106, USA}

\author[0000-0002-5258-6846]{Eric Gaidos}
\affiliation{Department of Earth Sciences, University of Hawai’i at Mänoa, Honolulu, Hawai’i 96822, USA}

\author[0000-0001-6532-6755]{Quang H. Tran}
\altaffiliation{51 Pegasi b Fellow}
\affiliation{Department of Astronomy, Yale University, New Haven, CT 06511, USA}

\author[0000-0002-2696-2406]{Jingwen Zhang}
\affiliation{Department of Physics, University of California, Santa Barbara, Santa Barbara, CA 93106, USA}

\author[0000-0002-4290-6826]{Judah Van Zandt}
\affiliation{Department of Physics, University of California, Santa Barbara, Santa Barbara, CA 93106, USA}

\author[0009-0004-3843-5285]{Katie E. Painter}
\affiliation{Department of Astronomy, The University of Texas at Austin, Austin, TX 78712, USA}

\author[0000-0003-0967-2893]{Erik A. Petigura}
\affiliation{Department of Physics \& Astronomy, University of California Los Angeles, Los Angeles, CA 90095, USA;}

\author[0000-0002-0726-6480]{Darryl Z. Seligman}
\affiliation{Department of Physics and Astronomy, Michigan State University, East Lansing, MI 48824, USA}

\author[0000-0002-9464-8101]{Adina D. Feinstein}
\affiliation{Department of Physics and Astronomy, Michigan State University, East Lansing, MI 48824, USA}

\author[0000-0002-5741-3047]{David R. Ciardi}
\affiliation{Caltech/IPAC-NASA Exoplanet Science Institute Pasadena, CA, USA}

\author[0000-0003-2102-3159]{Rocio Kiman}
\affiliation{Department of Physics, University of California, Santa Barbara, Santa Barbara, CA 93106, USA}

\author[0000-0003-3504-5316]{Benjamin J. Fulton}
\affiliation{Department of Astronomy, California Institute of Technology, Pasadena, CA 91125, USA}

\author[0000-0002-0531-1073]{Howard Isaacson}
\affiliation{Department of Astronomy, University of California, Berkeley, CA 94720, USA}

\author[0000-0001-8638-0320]{Andrew W. Howard}
\affiliation{Department of Astronomy, California Institute of Technology, Pasadena, CA 91125, USA}

\author[0000-0001-6187-5941]{Stefan Dreizler}
\affiliation{Institut für Astrophysik, Georg-August-Universität, Friedrich-Hund-Platz 1, 37077 Göttingen, Germany}

\begin{abstract}

51 Pegasi harbors the first confirmed extrasolar planet orbiting a Sun-like star. Decades of continued radial velocity (RV) observations have since uncovered signatures of an additional distant companion in the system from a shallow radial acceleration. We present new constraints on the mass and separation of a potential outer companion based on a synthesis of RVs, absolute astrometry, and new high-contrast imaging. Our analysis combines 31~years of new and previously published RV measurements from the OHP/ELODIE, Lick/Hamilton, Keck/HIRES, and APF/Levy spectrographs; a $\sim$25-year baseline of absolute astrometry from  Hipparcos and Gaia; and deep imaging from Keck/NIRC2 and HST/WFPC2. We find evidence for curvature in the RVs, which when combined with non-detections from imaging and astrometry point to a super-Jupiter at $\simeq$15--100~AU or brown dwarf companion at $\approx$20--170~AU.  However, the inferred radial acceleration of the host star is driven primarily by the Lick/Hamilton dataset and its slope is consistent with long-term instrument drift, calling into question the nature of the long-period signal. If an outer companion is present, it could explain the origin of the inner hot Jupiter if 51 Peg b arrived at its current location through high-eccentricity migration. On the other hand, if the signal is spurious, the exceptional baseline rules out Jovian planets within $\sim$10~AU and most brown dwarfs within several tens of AU, implying that the system is devoid of massive companions.  Continued RV and astrometric monitoring together with high-contrast imaging can be used to distinguish these scenarios.

\end{abstract}
\section{Introduction}\label{sec:intro}
Throughout the 1980s and 1990s, several teams launched the first pioneering searches for planetary companions to nearby stars using increasingly precise radial velocities (RVs; e.g. \citealt{CampbellWalker1979}; \citealt{CampbellWalker1988}; \citealt{Latham1989}; \citealt{MarcyBenitz1989}; \citealt{Walker1992}; \citealt{HatzesCochran1993}). Early results included a close-in massive planet candidate around HD 114762 (\citealt{Latham1989})\footnote{The short-period companion to HD 114762 B has since been shown to be a low-mass star on a nearly face-on orbit (\citealt{Kiefer2019}; \citealt{Winn2022}; \citealt{Holl2023}).} and a cold Jupiter in the $\gamma$ Cep binary system (\citealt{CampbellWalker1988}), although the existence of this planet was subsequently debated for many years (\citealt{Walker1992}; \citealt{Hatzes2003}). Soon after, \citet{MayorQueloz1995} presented the discovery of the first extrasolar planet orbiting the nearby, single main-sequence star 51 Pegasi with the ELODIE spectrograph (\citealt{Baranne1996}) mounted on the 1.9-m Observatoire de Haute-Provence telescope. 
\citet{MarcyButler1997} and \citet{HatzesCochran1998} independently conducted follow-up observations of the host star using the Hamilton echelle spectrograph (\citealt{Vogt1987}) at Lick Observatory and the coudé echelle spectrograph (\citealt{Tull1995}) on the 2.7-m Harlan J. Smith Telescope at McDonald Observatory.
Both studies confirmed the planet-like signal originally reported by \citet{MayorQueloz1995}. 

Prior to the discovery of 51 Peg b, theories of planet formation and planetary architectures were guided exclusively by the Solar System with inner small rocky planets segregated from the distant gas and ice giants. 
With an orbital period of $P=$ 4.23~d and a minimum mass of $M_p \textrm{sin}i$ = 0.47 $M_\mathrm{Jup}$, 51 Peg b immediately demonstrated that planets can take on a  greater diversity of orbital architectures than was appreciated from our limited view in the Solar System. It is now clear that these hot Jupiters with orbital periods $\lesssim$10~d are rare, with intrinsic occurrence rates of about 1$\%$ around single Sun-like stars (\citealt{Marcy2005}; \citealt{Mayor2011}, \citealt{Wright2012}, \citealt{BeleznayKunimoto2022}).  There is structure in the period distribution, with a ``few-day pileup'' of 3--5~d that may be a sign of halted inward migration (\citealt{Udry2003}; \citealt{Butler2006}; \citealt{YeeWinn2023}).  Hot Jupiters are more prevalent around both high-metallicity stars (\citealt{Guo2017}, \citealt{Petigura2018}) and high-mass stars (\citealt{Gan2022}; \citealt{Bryant2023}), although there are signs of a turnover at $\approx$1 $M_{\odot}$ beyond which the occurrence rates appear to be lower (\citealt{BeleznayKunimoto2022}).

Discoveries of the first hot Jupiters led to extensive theoretical efforts to explain the origins of this rare class of planets (see \citealt{Dawson2018}; \citealt{Fortney2021}). In particular, three prominent formation and migration channels have been proposed.  One possibility is that hot Jupiters form at larger orbital distances and undergo inward migration through early interactions with the protoplanetary disk (\citealt{GoldreichTremaine1980}; \citealt{LinBodenheimer1996}; \citealt{KleyNelson2012}). 
Alternatively, when eccentricities are excited and periastron distances shrink, tidal friction can dissipate orbital energy and circularize the planet's orbit. Giant planets may be driven into a high-$e$ state as a result of von Zeipel-Lidov-Kozai (ZLK) oscillations with an outer companion, in which eccentricity and inclination vary over long timescales (\citealt{Kozai1962}; \citealt{Lidov1962}; \citealt{Eggleton2001}; \citealt{FabryckyTremaine2007}; \citealt{Wu2007}), or through planet-planet scattering which can excite eccentricities especially in the case where the inner companion has a lower mass (\citealt{rasioford1996}; \citealt{chatterjee2008}). 
Finally, some hot Jupiters may have formed \emph{in situ} if favorable conditions related to inner disk surface density and core formation timescale are met (\citealt{Batygin2016}; \citealt{Boley2016}). 

Searches for outer companions to stars hosting hot Jupiters can help discern these pathways. An RV trend was evident in the residuals of the original ELODIE RV measurements of 51 Peg after removing the signal from 51 Peg b, which \citet{MayorQueloz1995} interpreted as evidence for an additional long-period exterior companion. A shallower long-term radial acceleration was also highlighted by \citet{Butler2006} based on Hamilton RVs spanning 6 years, further suggesting the presence of an outer giant planet,  brown dwarf, or distant stellar companion. \citet{Birkby2017} analyzed a compilation of archival RV measurements taken over 20 years from ELODIE, Hamilton, HARPS, and HIRES; they found differences in the inferred slope of the trend with various datasets---a sign that there may be a turnover in the RVs during this period.
A long-period signal was also noted by \citet{Rosenthal2021} for 51 Peg based on Hamilton, HIRES, and APF measurements taken over a 26-year baseline.

\begin{deluxetable}{cccc}
\renewcommand\arraystretch{0.9}
\tabletypesize{\small}
\setlength{\tabcolsep}{0.25cm}
\tablewidth{0pt}
\tablecolumns{4}
\tablecaption{New APF and HIRES Relative Radial Velocities of 51 Peg\label{tab:51_Pegasi_RVs}}
\tablehead{
    \colhead{BJD} &
    \colhead{RV} &
    \colhead{$\sigma_\mathrm{RV}$} &
    \colhead{Instrument} \\
    & \colhead{(m s$^{-1}$)} & \colhead{(m s$^{-1}$)} &
}
\startdata
2459966.60283 & -51.91 &  3.10 & APF \\
2459968.59859 & 54.66 &  4.32 & APF \\
2459971.59723 & -10.33 &  3.96 & APF \\
2459976.60235 & 46.57 &  3.03 & APF \\
2460128.99502 & 49.53 &  3.97 & APF \\
2460131.95315 & -37.07 &  2.90 & APF \\
2460133.98319 & 48.30 &  3.68 & APF \\
2460135.99968 & -53.43 &  3.77 & APF \\
2460214.75186 & 27.25 &  3.48 & APF \\
2460233.72405 & -23.32 &  3.55 & APF \\
2460238.67556 & 30.85 &  3.34 & APF \\
2460252.65951 & 31.93 &  3.12 & APF \\
2460563.82532 & -17.29 &  3.10 & APF \\
2460827.97072 & 36.23 &  2.19 & APF \\
2459004.12133 & 46.68 &  1.07 & HIRES \\
2459047.04229 & 14.49 &  1.25 & HIRES \\
2459072.85525 & -19.29 &  1.21 & HIRES \\
2459155.85953 & 38.00 &  1.24 & HIRES \\
2459382.11558 & -45.60 &  1.16 & HIRES \\
2459398.07252 & 22.98 &  1.08 & HIRES \\
2459417.91345 & 21.22 &  1.27 & HIRES \\
2459418.04586 & 36.58 &  1.42 & HIRES \\
2459418.05205 & 36.19 &  1.08 & HIRES \\
2459592.69364 & 27.93 &  1.21 & HIRES \\
2459947.68791 & 42.06 &  1.17 & HIRES \\
2459953.69176 & -68.28 &  1.27 & HIRES \\
2460148.14054 & -56.98 &  1.23 & HIRES \\
\enddata
\tablecomments{These measurements should be considered as an extension of the APF and HIRES RVs from \citet{Rosenthal2021}.}
\end{deluxetable}

Here we present new constraints on the mass, separation, and luminosity of a potential second companion in the 51 Peg system. We synthesize 891 RVs taken over 31 years from four spectrographs, absolute astrometry of 51 Peg from Hipparcos and Gaia, and new high-contrast imaging from Keck and the \textit{Hubble Space Telescope} (\emph{HST}) to clarify the nature of the potential distant companion, search for new low-mass planets in the 51 Peg system, and establish limits on additional long-period substellar companions.
 
 The remainder of this paper is organized as follows. In Section \ref{sec:Observations} we provide an overview of the various observations of the host star considered in this study and details of the orbit fitting procedure. In Section \ref{sec:Results} we contextualize our results and our interpretation of the longer-period companion. The impact of stellar activity and the possible connection to the migration of 51 Peg b are  discussed in Section \ref{sec:Discussion}. Finally, we summarize our conclusions in Section \ref{sec:Conclusion}.

\section{Observations and Time Series Analysis}\label{sec:Observations}

\begin{figure*}
\begin{center}
\includegraphics[width=\textwidth,height=0.55\textheight,keepaspectratio]{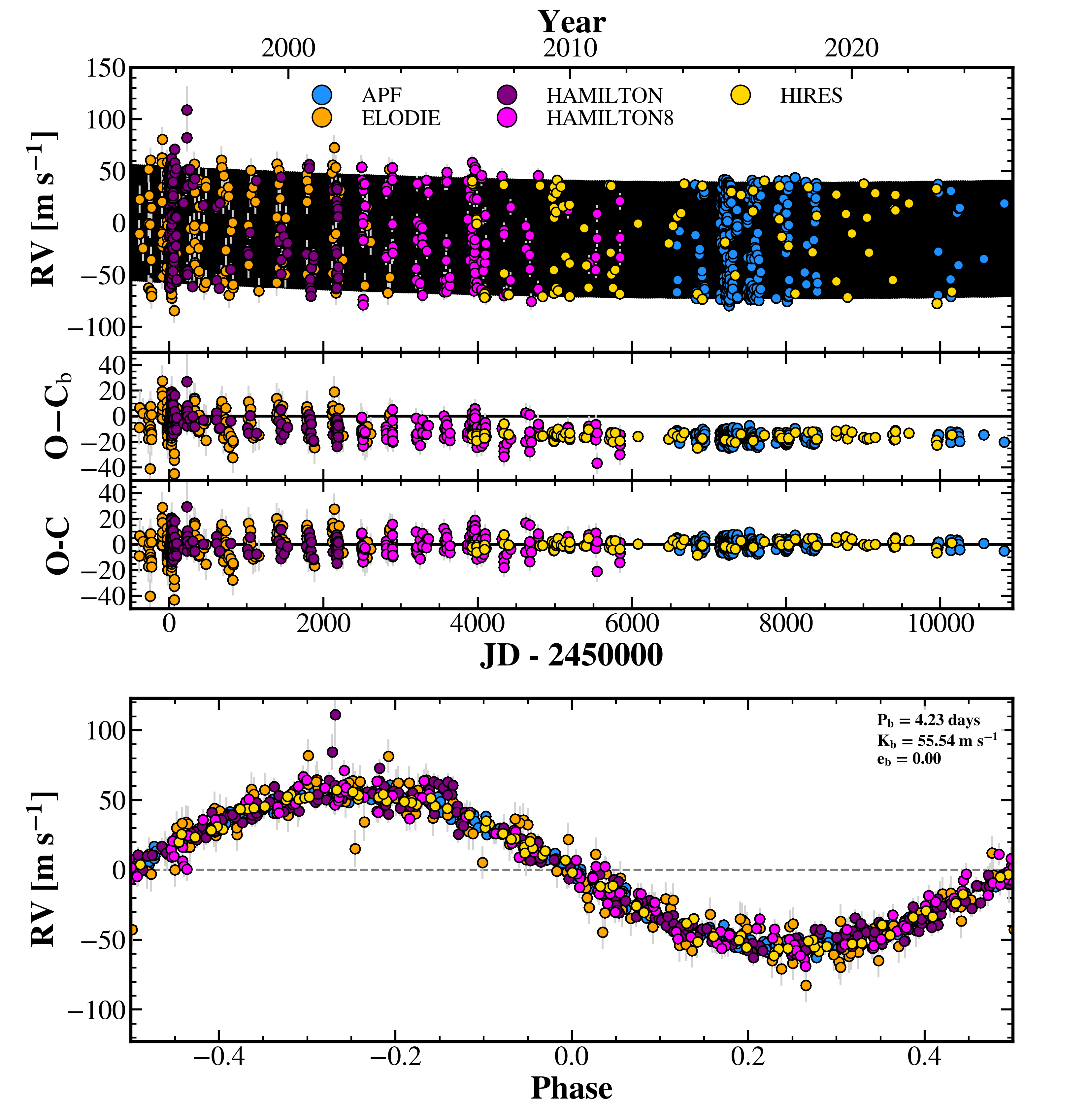}
\caption{Best-fit model to RVs of 51 Peg.  The preferred solution includes both the strong signal from 51 Peg b and long-term curvature---a change in radial acceleration. The top panel shows the full 31-year RV time series combining RVs from APF, ELODIE, Hamilton, and HIRES. Underneath are two sets of residuals: those after subtracting the inner hot Jupiter with the acceleration signal retained, and those after subtracting 51 Peg b and the long-term acceleration. The phase-folded orbit of the inner planet, 51 Peg b, is displayed in the bottom panel. }
\label{fig:combined_rv_plot}
\end{center}
\end{figure*}

\subsection{New APF/Levy and Keck/HIRES RVs}\label{sec:new_RVs}
We include in our analysis 27 new previously unpublished precision RVs, 13 from Keck/HIRES and 14 from APF/Levy, measured between early 2020 and mid 2025. The APF RVs were acquired as part of the APF-50 survey of Sun-like stars \citep{FultonThesis2017} and have a median photon-limited uncertainty of 3.4 m $\text{s}^{-1}$. The HIRES RVs have a median uncertainty of 1.2 m $\text{s}^{-1}$. We performed the observations and data reduction for both data sets according to the standard procedures of the California Planet Search
\citep{Howard2010_I2technique, Rosenthal2021}. Briefly, our observational method involves acquiring stellar spectra with a warmed cell of molecular iodine mounted directly in front of the spectrograph entrance slit. The dense set of absorption lines imprinted by the iodine serves as a wavelength reference from which to measure slight shifts in the stellar spectrum. The RV for each spectrum is fit as a free parameter in a forward model of the combined iodine-star spectrum. These new observations match the reference frame of \cite{Rosenthal2021}, and extend the RV baseline of 51 Peg from 26 to 31 years (see Table \ref{tab:51_Pegasi_RVs}).

\subsection{RV Analysis}\label{sec:RV_Analysis}
 We combine independent RV datasets from several instruments that have monitored 51 Peg including the ELODIE spectrograph on the OHP 1.9-m telescope (153 RVs; \citealt{Baranne1996}), Hamilton on the Lick/Shane 3-m telescope (354 RVs; \citealt{Vogt1987}), HIRES on the Keck-I 10-m telescope (87 RVs; \citealt{Vogt1994}), and APF on the Lick/Levy 2.4-m telescope (297 RVs; \citealt{Vogt2014}).\footnote{A total of 308 RV measurements were obtained with the HARPS spectrograph between 2013 and 2018. However, the two HARPS datasets are separated by approximately five years, lack intermediate observations, and are split by a fiber upgrade that introduced an RV offset. We therefore excluded the HARPS RVs from this analysis.} The ELODIE RVs are compiled from \citet{Naef2004}, the Hamilton RVs from \citet{Fischer2014}, and the HIRES and APF RVs from \citet{FultonThesis2017}, \citet{Rosenthal2021}, as well as the new observations described in Section~\ref{sec:new_RVs}. Combined, these datasets establish a 31-year baseline of 891 high-precision RVs of 51 Peg between 1994 and 2025.

\citet{Butler2006} reported a linear trend of --1.64 m $\text{s}^{-1}$ yr$^{-1}$ for 51 Peg based on 6 years of RV measurements from the Hamilton spectrograph. Long-term RV trends can be caused by distant giant planets, brown dwarfs, or low-mass stellar companions, revealed through long observational baselines (e.g., \citealt{Crepp2014}; \citealt{Lagrange2019}; \citealt{Bowler2021A}; \citealt{Bowler2021B}; \citealt{Dalal2021}; \citealt{Barbato2023}; \citealt{Matthews2024}; \citealt{BardalezBalmer2025}). However, stellar magnetic activity cycles can also mimic low-amplitude RV accelerations through periodic elevated and reduced spot covering fractions.  Similar trends can be caused by instrumental drifts as a result of slowly varying environmental changes. Here we attribute the signal to the reflex motion of a distant companion, and in Sections~\ref{sec:hamilton_trend_artifact} and \ref{sec:Stellar_Activity} we explore the possibility of an instrumental drift and cyclical stellar activity.

We use the \texttt{RadVel} package (\citealt{Fulton2018}) to model the full RV time series data and assess if there is evidence for a distant companion in the combined dataset. Several cases are considered (see below), with model parameters determined from Markov-chain Monte Carlo (MCMC) posterior sampling using an affine-invariant algorithm (\citealt{Foreman-Mackey2013}). Each MCMC run contains 70 walkers and 10$^4$ steps. We utilize the built-in Gelman-Rubin (\citealt{GelmanRubin1992}; GR) statistic in \texttt{RadVel} (which imposes a minimum GR value of 1.03) coupled with trace and corner plots to assess convergence (\citealt{Foreman-Mackey2016}; \citealt{HoggForeman-Mackey2018}). 

\setlength{\tabcolsep}{8pt}
\begin{deluxetable*}{lccc}
\renewcommand{\arraystretch}{1.2}
\tabletypesize{\scriptsize}
\tablecaption{Priors and Posteriors for 51 Pegasi system. \label{tab:model_params}}
\tablehead{
\colhead{Parameter} & 
\colhead{Adopted Prior} & 
\colhead{Posterior Value} 
}
\startdata
\textbf{Fitted Parameters} & &  \\
$T_c$ (BJD$-2450000$)&   $\mathcal{U}$[$\tau$-$\frac{1}{2}$$P_{0}$, $\tau$+$\frac{1}{2}$$P_{0}$] &  6326.9333 (0.0002, 0.0002) \\
$P$ (days) &  $\mathcal{U}$[1, 10] &  4.2307916 (0.0000003, 0.0000003) & \\
$K$ (m s$^{-1}$) & $\mathcal{J}$[1, 1000] & 55.53 (0.01, 0.01)  \\
$\sqrt{e}\sin\omega$ & Fixed & 0\\
$\sqrt{e}\cos\omega$ & Fixed & 0\\
\hline
\textbf{Derived Parameters} & &  \\
$m_p \sin i$ $M_\mathrm{Jup}$ &  $\cdots$ & 0.4621 (0.0001, 0.0001) \\
$e$ &  Fixed & 0 \\
$\omega$ (deg) &  Fixed & 0 \\
$a$ (AU) &  $\cdots$ &  0.052344942 (0.000000002, 0.000000002) \\
\hline
\textbf{Additional Parameters} & &  \\
$\dot{\gamma}$ (m $\text{s}^{-1} \text{d}^{-1}$)&  $\mathcal{U}$[--2.75, 2.75] & -0.00403 (0.00004, 0.00004) \\
$\ddot{\gamma}$ (m $\text{s}^{-1} \text{d}^{-2}$)&  $\mathcal{U}$[--0.0075, 0.0075] & 0.000000240 (0.000000002, 0.000000002) \\
RV jitter (APF) (m s$^{-1}$) &  $\mathcal{U}$[0, 100] & 1.99 (0.01, 0.01) \\
RV jitter (ELODIE) (m s$^{-1}$) &  $\mathcal{U}$[0, 100] & 9.39 (0.09, 0.05) \\
RV jitter (Hamilton) (m s$^{-1}$) &  $\mathcal{U}$[0, 100] & 6.36 (0.03, 0.02) \\
RV jitter (Hamilton8) (m s$^{-1}$) &  $\mathcal{U}$[0, 100] & 7.18 (0.04, 0.05) \\
RV jitter (HIRES) (m s$^{-1}$) & $\mathcal{U}$[0, 100] & 2.68 (0.02, 0.03) \\
\enddata
\tablecomments{Best-fit parameters for 51 Peg~b and constraints on its distant companion from RVs. The MAP value is reported along with the upper and lower bounds of the 68.3$\%$ highest density interval regions of the posterior probability distributions. 
Instrument offsets are calculated analytically: $\gamma_{\mathrm{APF}}$ = +20.40 m s$^{-1}$, $\gamma_{\mathrm{ELODIE}}$ = --33247.50 m s$^{-1}$, $\gamma_{\mathrm{Hamilton}}$ = +24.13 m s$^{-1}$, $\gamma_{\mathrm{Hamilton8}}$ = +5.74 m s$^{-1}$, and $\gamma_{\mathrm{HIRES}}$ = +11.81 m s$^{-1}$. }
\tablecomments{$\mathcal{U}$ and $\mathcal{J}$ represent linear uniform and Jeffreys priors, respectively and $\tau$ refers to an arbitrary reference value.}
\end{deluxetable*}

\begin{deluxetable*}{lcccccccccccl}
\renewcommand\arraystretch{0.9}
\tabletypesize{\footnotesize}
\setlength{\tabcolsep} {.1cm}
\tablewidth{0pt}
\tablecolumns{5}
\tablecaption{Orbit fits to individual and combinations of datasets. \label{tab:spectrographs_bics}}
\tablehead{\colhead{Instrument}  & \colhead{$\lambda$} & \colhead{$N_\mathrm{inst}$} & \colhead{$\#$ RVs} & \colhead{Start Date} & \colhead{End Date} & \colhead{$\Delta$$t$} &\colhead{$\Delta$$\mathrm{BIC_{K}}$} & \colhead{$\Delta$$\mathrm{BIC_{L-K}}$} & \colhead{$\Delta$$\mathrm{BIC_{Q-K}}$} &\colhead{Reference}  \\
 &\colhead{(nm)} & & & \colhead{(UTC)}& \colhead{(UTC)}& \colhead{(years)}& & & & 
}
\startdata
APF & 374--970 & 1& 297 & 2013--10--16 & 2025--06--01 & 11.6 & \textbf{0.0} &5.6& 8.3& 1,2,3 \\
ELODIE  & 390--681 & 1& 153 & 1994--09--15 & 2003--09--05 & 9.0 & \textbf{0.0} & 4.90 & 8.72 & 4,5,6 \\
Hamilton13+6 $\&$ Hamilton8 & 340--1100 & 2 & 354 & 1995--10--12 & 2011--10--12 & 16.0 & 0.0 & \textbf{--35.2}& --41.3 & 7,8 \\
HIRES & 300--1100 & 1& 87 & 2006--07--10 & 2023--07--22 & 17.0 & \textbf{0.0} & 3.9& 7.2 & 3, 9 \\
Full dataset & $\cdots$ & 5 & 891 & 1994--09--15 & 2025--06--01 & 30.7 & 0.0 & --0.20& \textbf{--25.8}& $\cdots$ \\
Hamilton$_{\text{combined}}$\tablenotemark{a} & 340--1100  & 1& 354 &  1995--10--12 & 2011--10--12 & 16.0 & 0.0 & \textbf{--467.8} & --474.7 & $\cdots$ \\
APF $\&$ HIRES & $\cdots$ & 2& 384 & 2006--07--10 & 2025--06--01 & 18.9 & \textbf{0.0} & 6.4 & 10.0 & $\cdots$ \\
ELODIE $\&$ Hamilton$_{\text{subset}}$\tablenotemark{b} & $\cdots$ & 2& 403 & 1994--09--15& 2003--09--08 & 9.0 & 0.0 & --141.5 & \textbf{--174.5} & $\cdots$ \\
ELODIE $\&$ APF $\&$ HIRES & $\cdots$ & 3 & 537 & 1994--09--15& 2025--06--01 & 30.7 & \textbf{0.0} & 5.43 & 10.59 & $\cdots$ \\
\enddata
\tablenotetext{a}{The Hamilton$_{\text{combined}}$ RV dataset represents the Hamilton measurements treated as one combined instrument. The relative offsets for the Hamilton13, Hamilton8, and Hamilton6 measurements are adopted from \citet{Fischer2014}, which were calibrated from RV standards.}
\tablenotetext{b}{Hamilton$_{\text{subset}}$ represents a subset of the Hamilton RV dataset matched to the temporal baseline of the ELODIE RVs. There are 249 RVs spanning 1995--10--12 to 2003--09--08.}
\tablecomments{\small Telescopes and instruments used to discover and characterize 51 Peg b.  We present $\Delta$BIC values for all three models with the Keplerian serving as a reference model ($\Delta$$\mathrm{BIC_{K}} \equiv 0$): Keplerian+Linear trend versus Keplerian-only ($\Delta$$\mathrm{BIC_{L-K}}$), or Keplerian+Quadratic trend versus Keplerian-only ($\Delta$$\mathrm{BIC_{Q-K}}$). The number of free parameters for each system can be computed as  $3 + N_\mathrm{inst}$ + acceleration terms used in the model, where a linear trend corresponds to one acceleration term and curvature is two.}
\tablerefs{(1) \citet{Vogt2014}; (2) \citet{FultonThesis2017}; (3) \citet{Rosenthal2021}; (4) \citet{MayorQueloz1995}; (5) \citet{Baranne1996}; (6) \citet{Naef2004}; (7) \citet{Vogt1987}; (8) \citet{Fischer2014}; (9) \citet{Vogt1994}.}
\end{deluxetable*}

We choose to perform the orbit fits with the standard basis for efficient posterior sampling using orbital period $P$, time of inferior conjunction $T_{C}$, semi-amplitude $K$, and parameterized forms of eccentricity ($e$) and argument of periastron ($\omega$), $\sqrt{e}$ $\cos \omega$ and $\sqrt{e}$ $\sin \omega$. We also fit for jitter ($\sigma_i$, with one per instrument $i$), instrumental offsets ($\gamma_i$, with one per instrument $i$), and, for models that include a distant companion, a linear acceleration ($\dot{\gamma}$) and quadratic acceleration term ($\ddot{\gamma}$). 
Uniform priors are adopted with bounds on the period, time of conjunction, jitter, linear acceleration, and quadratic acceleration to enforce physically meaningful constraints. For the RV semi-amplitude ($K$), a bounded Jeffreys prior is used to enforce positive values and avoid bias toward small amplitudes. To aid with convergence and focus on the acceleration of the host star rather than the orbit of the inner companion, we assume 51 Peg b has been completely circularized and fix the eccentricity and argument of periastron by setting $\sqrt{e} \cos \omega$ and $\sqrt{e} \sin \omega$ to 0 in all orbit fits. This is consistent with previous fits that found an upper limit on the eccentricity of 0.0042$^{+0.0046}_{-0.0030}$ (\citealt{MayorQueloz1995}; \citealt{Naef2004}; \citealt{Rosenthal2021}).  A summary of the adopted priors can be found in Table~\ref{tab:model_params}.

We perform three orbital fits for comparison: 
(1) a single-planet Keplerian orbit (51 Peg b) with no RV trend in the model (8 free model parameters), (2) a single planet plus a linear trend (a radial acceleration; 9 free parameters), and (3) a single planet with both linear and curvature terms (changes in radial acceleration; 10 free parameters). We compare the Keplerian model against the linear trend and curvature (quadratic) models using the Bayesian Information Criterion (BIC; \citealt{Schwarz1978}; \citealt{Claeskens2016}), which balances model complexity with goodness of fit. Following the interpretation of $\Delta$BIC values from \citet{Trotta2007}, we then select the statistically favored fit with the lowest BIC value if the $\Delta$BIC for that model is $>$10 (a sign of ``strong evidence'') compared to the less complex, next-best fit model.\footnote{A $\Delta$BIC value between 2 and 5 suggests positive evidence, while a $\Delta$BIC value between 5 and 10 indicates moderate evidence in favor of the model with the lower BIC.} In this way we favor the simplest model and only invoke a more complex model if it is sufficiently justified by an improvement in the fit.  The orbit fit with curvature is statistically preferred over a single Keplerian fit ($\Delta$BIC = 25.8) and a Keplerian modeled with a linear trend ($\Delta$BIC = 25.6). In addition to fitting a Keplerian model, as well as Keplerian models including a linear trend and a second-order acceleration term (curvature), we also considered models incorporating third-order and fourth-order acceleration terms. We find that the model including curvature provides the best fit to the time-series data. The radial accelerations, $\dot{\gamma}$ and $\ddot{\gamma}$, have values of --1.47$^{+0.01}_{-0.01}$ m $\text{s}^{-1}$ yr$^{-1}$ and 0.0300$^{+0.0003}_{-0.0003}$ m $\text{s}^{-1}$ yr$^{-2}$, respectively.  Results of the orbit fits can be found in Table~\ref{tab:model_params}, Figure~\ref{fig:combined_rv_plot}, and Appendix~\ref{sec:Additional_Statistical_Tests}.  Note that $\dot{\gamma}$ and $\ddot{\gamma}$ have the strongest covariance among all parameters in the orbit fit. Additional RV fits can be found in Appendix \ref{sec:Additional_RV_Observations}.

\subsection{Results from Instrument Subsets}\label{sec:Previous Results}

\citet{Birkby2017} provided the most thorough assessment of trends identified in various RV datasets targeting 51 Peg to date.  They performed individual fits to each of the spectrographs, all of which differ in timing and duration of the RV monitoring, and found that the resulting slopes are often mutually inconsistent. They noted that the varying inferred RV trends from their broader compilation of RVs up to 2014 may have probed the turnover point of an RV curve for a long-period companion. However, they did not analyze the nature of this possible distant object any further. Our analysis includes an additional 334 RVs spanning 11 more years from \citet{FultonThesis2017}, \citealt{Rosenthal2021}, and this work with which we aim to further explore this possibility. In addition to modeling the combined RV time series (Section~\ref{sec:RV_Analysis}), here we also model individual datasets to compare with results from previous efforts.

\subsubsection{ELODIE}
\citet{Birkby2017} modeled the ELODIE RVs (spanning 1994--2003) with a planet signal-plus-linear trend and found a small value of --0.15$^{+0.37}_{-0.40}$ m $\text{s}^{-1}$ yr$^{-1}$, which is consistent with no long-term trend. When we model these same ELODIE RVs with a planet-plus-linear trend, this yields a similar value of --0.14$^{+0.03}_{-0.03}$ m $\text{s}^{-1}$ yr$^{-1}$. However, we find that a Keplerian model without an acceleration is statistically preferred over models that include either a linear trend or curvature. For the ELODIE dataset alone, we therefore do not find signs of a long-term signal.

\subsubsection{Hamilton}\label{sec:hamilton_subset}
The Hamilton Spectrograph experienced several CCD upgrades, with a naming convention related to the associated dewar in which each detector was mounted--Hamilton13, Hamilton6, and Hamilton8 (see \citealt{Fischer2014}).  
Based on RV standards with more than 100 observations taken between 1995 and 2011, \citet{Fischer2014} note that the inferred offset between the Hamilton13 and Hamilton6 RVs is smaller than the typical RV measurement errors.  However, the Hamilton8 zero-point offset was found to be 13.1 m s$^{-1}$. Although this offset is applied to the RVs reported in \citet{Fischer2014}, for the global fit we chose to model the Hamilton8 RVs as an independent spectrograph (which we refer to as ``Hamilton8,'' while the Hamilton13 and Hamilton6 RVs are called ``Hamilton13+6''; see Figure \ref{fig:combined_rv_plot}) to adopt the most conservative assumptions.

\citet{Birkby2017} opted to model the Hamilton RVs (spanning 1995--2011) as three independent datasets. They found linear trends of --1.64$^{+1.17}_{-1.10}$ m $\text{s}^{-1}$ yr$^{-1}$, 0.029$^{+0.28}_{-0.29}$ m $\text{s}^{-1}$ yr$^{-1}$, and --0.58$^{+0.84}_{-0.88}$ m $\text{s}^{-1}$ yr$^{-1}$ for Hamilton13, Hamilton6, and Hamilton8, respectively. \citet{Butler2006} modeled the Hamilton RVs collected until 2006 and reported a linear trend of --1.64 m $\text{s}^{-1}$ yr$^{-1}$. Recently, \citet{Pena2025} modeled the Hamilton RV time series of 51 Peg reported in \citet{Butler2006} to test their MCMC parallel tempering RV orbit fitting code, \texttt{EMPEROR}; they also confirm a similar linear trend in the system. 

To compare with previous results, we carry out two fits of the Hamilton-only dataset. We first model all of the Hamilton RVs together as one spectrograph (``Hamilton$_\mathrm{combined}$'') with a planet-plus-linear trend and find a similar slope as identified by \citet[for Hamilton13 and Hamilton6]{Butler2006} and \citet[for Hamilton13]{Birkby2017}, albeit with tighter uncertainties: --1.583$^{+0.004}_{-0.004}$ m $\text{s}^{-1}$ yr$^{-1}$. When we allow the Hamilton8 subset of RVs to float and fit for a separate instrument offset, we find that the Hamilton RVs continue to statistically favor a linear trend of --1.35$^{+0.01}_{-0.01}$ m $\text{s}^{-1}$ yr$^{-1}$  with a $\Delta$BIC of 35 as compared to the Keplerian-only model. 

\subsubsection{ELODIE and Hamilton} 
We analyze the ELODIE data set together with a subset of the Hamilton data (``Hamilton$_\mathrm{subset}$''), matched in baseline, to test for trends in the two earliest data sets. The combination of these two datasets strongly favors a curvature model with values of $\dot{\gamma}$ = 3.46$^{+0.06}_{-0.07}$ m $\text{s}^{-1}$ yr$^{-1}$ and --0.74$^{+0.01}_{-0.01}$ m $\text{s}^{-1}$ yr$^{-2}$. The curvature model is preferred over both a model that incorporates a linear trend ($\Delta$BIC = 33) and a Keplerian-only model ($\Delta$BIC = 175).

\subsubsection{HIRES and APF}

Based on 43 HIRES RV observations acquired over eight years, \citet{Bryan2016} reported a linear trend of --0.42$^{+0.20}_{-0.20}$ m $\text{s}^{-1}$ yr$^{-1}$. With an extended baseline of observations, \citet{Rosenthal2021} used \texttt{RVSearch} to search for additional long-period companions in this system using the combined HIRES, APF, and Hamilton data.  They report a potential signal at 43$^{+41}_{-21}$ AU and attribute it to ``annual and/or instrumental systematics'' which may arise from dewar offsets in the Hamilton data, among other sources. (As discussed previously, we address the dewar offsets associated with the Hamilton data following recommendations in \citealt{Fischer2014}.)

We further perform independent tests for long-term trends with the HIRES and APF data presented in \citet{FultonThesis2017}, \citet{Rosenthal2021}, and this work. We first model all of the HIRES RV data together and find that the Keplerian model without an acceleration is favored over models that include a linear trend or curvature. Although not statistically favored, the next best fit model includes a shallow linear trend with a slope of --0.05$^{+0.01}_{-0.01}$ m $\text{s}^{-1}$ yr$^{-1}$. Following the same approach, the APF RVs similarly favor a Keplerian model without a linear trend or curvature. The next-best-fitting model, consisting of a Keplerian plus a linear trend, yields a slope of $-0.04^{+0.01}_{-0.01}\,\mathrm{m\,s^{-1}\,yr^{-1}}$.
Finally, a joint fit of the HIRES and APF datasets indicates the Keplerian-only model is justified.  We conclude that, both alone or together, the HIRES and APF RVs do not show convincing signs of an outer companion.

\subsubsection{Elodie, HIRES, and APF}
Here, we combine the ELODIE, HIRES, and APF measurements to further test whether the radial acceleration is present in these datasets, or instead if it is solely dependent on the Hamilton RVs. We find that the Keplerian-only model is preferred over both a model that incorporates a linear trend ($\Delta$BIC = 5) and curvature ($\Delta$BIC = 10). This suggests that the inferred shallow reflex motion of 51~Peg is primarily driven by the Hamilton dataset; there is no significant evidence of a companion when the Lick RVs are excluded.

\subsubsection{Summary of the RV Tests}
To summarize, we detect a statistically significant trend in the Hamilton dataset. A trend is favored for the Hamilton$_\mathrm{combined}$ test, in which the Hamilton RVs are treated as one instrument, and when the Hamilton8 RVs are treated as a separate instrument. Curvature is preferred both for the ELODIE + Hamilton$_\mathrm{subset}$ RVs and for the full dataset. 

While a Keplerian-only model is favored in the ELODIE, HIRES, and APF datasets, shallow linear radial accelerations cannot be conclusively ruled out as all have $\Delta$BIC values $<$6 between the Keplerian-only and Keplerian+Linear trend fits (see Table \ref{tab:spectrographs_bics}). We also find consistent values compared to published fits when using the same model and datasets adopted in previous studies. Taken at face value, this reinforces the existence of low-amplitude, evolving long-term modulations in the 51 Peg system, which in Section~\ref{sec:RV_Analysis} we found is best modeled collectively with a curvature solution.

However, we also found that the inferred long-term signal in the RV timeseries data is dependent on the Hamilton dataset. Without this, the Elodie, HIRES, and APF RVs are most consistent with a single planet---51 Peg b.  As discussed  in Section~\ref{sec:hamilton_subset}, we account for offsets in the Hamilton data by conservatively modeling the ``Hamilton13+6'' and ``Hamilton8'' datasets separately, so the trend is unlikely to be caused by these intra-instrument zero points. 
In Section~\ref{sec:hamilton_trend_artifact}, we explore whether the Hamilton signal can be caused by low-amplitude RV instrumental drifts by comparing RV standard stars that are common to both long-baseline Hamilton and HIRES datasets.

\subsubsection{Secular Acceleration}
At a distance of 15.51 pc, 51 Peg is close enough that the radial and tangential components of its three-dimensional velocity vector can produce a perspective (or “secular”) acceleration.  The impact is larger for closer stars and those with higher proper motions.  Perspective acceleration manifests as an apparent change in radial velocity that can mimic a long-term trend caused by a wide-separation companion (\citealt{Choi2013}). Following \citet{Zechmeister2009b}, this radial acceleration is

\begin{equation}
    \frac{d v_r(t)}{dt} = 22.98 \,\frac{\mathrm{m\,s^{-1}}}{\mathrm{yr}}\,
    \frac{((\mu_\alpha \cos \delta)^2 + \mu_\delta^2)\,\mathrm{yr}^2/\mathrm{arcsec}^2}{\pi/\mathrm{mas}} \, ,
\end{equation}

\noindent where $\mu_\alpha \cos \delta$ and $\mu_\delta$ are the right ascension and declination proper motion components, respectively, in units of $''$ yr$^{-1}$, and $\pi$ is the parallax angle in mas. 51 Peg has a secular acceleration of 0.0167 m $\text{s}^{-1} \text{yr}^{-1}$, or about 0.50 m $\text{s}^{-1}$ over 30 years. The inferred acceleration of $0.0167~\mathrm{m~s^{-1}~yr^{-1}}$ is significantly smaller than the 1--2 $\mathrm{m~s^{-1}~yr^{-1}}$ slope measured from Doppler monitoring. The trend therefore cannot be caused by this geometric effect.

\begin{figure*}
\begin{center}
\includegraphics[width=0.8\textwidth]{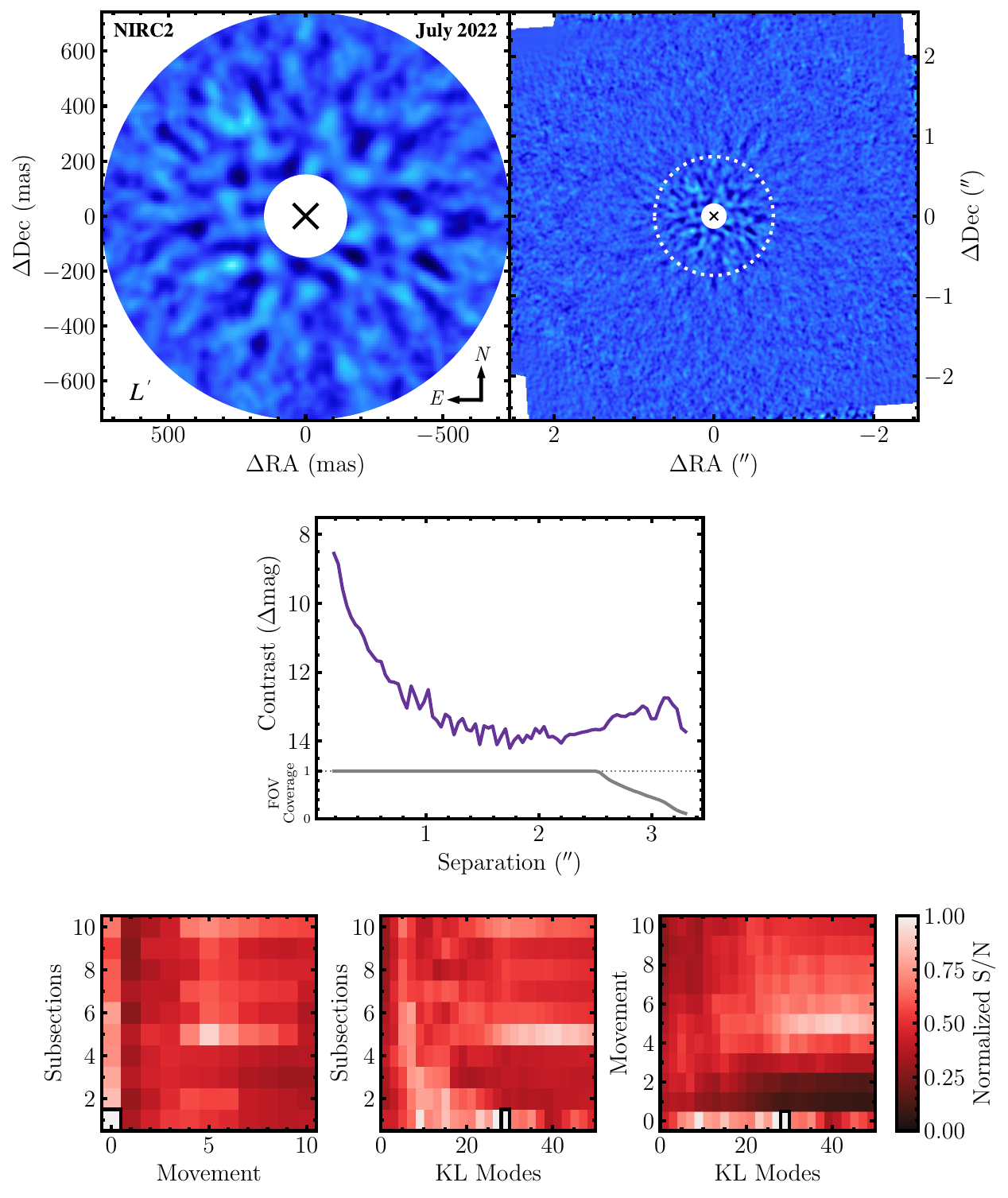}
\caption{\emph{Top:} Keck/NIRC2 $L'$ imaging of 51 Peg. The left panel shows a $750 \times 750 \, \mathrm{mas}$ cutout. The right panel shows the full frame image. Each image is oriented such that north is up and east is to the left. \emph{Center:} 5-$\sigma$ contrast curve (purple) and field of view coverage (grey). \emph{Bottom:} Recovered S/N of the injected source for optimizing \texttt{pyKLIP} parameters. Each panel shows the maximum recovered S/N across a grid of two parameters marginalized across the third parameter. The recovered S/N is normalized to the maximum S/N found through the optimization. The parameter combination that best recovers the injected source is highlighted in black.}
\label{fig:nirc2_imaging}
\end{center}
\end{figure*}

\subsection{Absolute Astrometry Analysis}\label{sec:Astrometry_Analysis}
Absolute astrometry provides an independent assessment of reflex motion in the plane of the sky caused by a distant companion.
For this study, we examine constraints on the acceleration of 51 Peg from the \textit{Hipparcos}-\textit{Gaia} EDR3 Catalog of Accelerations (HGCA; \citealt{Brandt2021}). The HGCA utilizes a 25-year baseline comparing proper motion measurements from \textit{Hipparcos}  (\citealt{vanLeeuwen2007}) and \textit{Gaia} EDR3 (\citealt{GaiaCollaboration2022}). It provides recalibrated measurements from the \textit{Hipparcos} ($\sim$1991) and \textit{Gaia} ($\sim$2016) missions together with a joint average proper motion from the difference in sky position between the two epochs. \citet{Kervella2019} carried out a similar analysis and reported proper motion anomalies as the difference between long-term \textit{Hipparcos}-\textit{Gaia} and short-term \textit{Gaia} proper motion vectors. Both catalogs have been used to directly image new long-period giant planets and brown dwarfs and constrain the orbits of known low-mass companions (e.g., \citealt{Brandt2019}; \citealt{Bowler2021A}; \citealt{Franson2023}; \citealt{DeRosa2023}; \citealt{Mesa2023}; \citealt{YitingLi2023}; \citealt{Matthews2024}; \citealt{An2025}; \citealt{BardalezBalmer2025}).

51 Peg exhibits changes in proper motion of $\Delta \mu_{\alpha, \mathrm{Gaia-HG}}$ = --0.26 $\pm$ 0.12 mas yr$^{-1}$ and $\Delta \mu_{\delta, \mathrm{Gaia-HG}}$ = 0.05 $\pm$ 0.10 mas yr$^{-1}$ from the HGCA, which translates to a tangential acceleration of 1.59 $\pm$ 0.68 m $\text{s}^{-1}$ yr$^{-1}$.  The $\chi^2$ value of a constant proper-motion model is 4.73 for 2 degrees of freedom, which corresponds to a significance level of 1.68$\sigma$ following the methodology detailed in \citet{Painter2025}.  Therefore, there is no evidence that 51 Peg exhibits a statistically significant astrometric acceleration in HGCA (EDR3). Note that the significance level using the DR2 version of HGCA (\citealt{Brandt2019}) is substantially higher: $\chi^2$ = 10.1 (2 degrees of freedom), or 2.7$\sigma$. \citet{Kervella2019} similarly found a proper motion anomaly of the stellar host between \textit{Gaia}~DR2 and \textit{Hipparcos} at the $3.04\sigma$ level, suggesting marginal evidence for a tangential acceleration. However, with the updated astrometric information in \textit{Gaia}~EDR3, \citet{Kervella2022} reported a tangential acceleration between \textit{Gaia}~DR3 and \textit{Hipparcos} at the $2.04\sigma$ level to which is more consistent with that derived from the HGCA (EDR3).

Nevertheless, the non-detection of an acceleration still provides useful information about the parameter space in which distant companions cannot reside.
Using equations 15 and 16 from \citet{Kervella2019}, and following the framework in \citet{Painter2025} for computing upper limit sensitivity curves, we are able to exclude the presence of various companions in mass and separation space (in this case at the 3$\sigma$ level) that would otherwise induce a measurable astrometric reflex motion on the host star.
These limits from astrometry can be considered orthogonal to the RV limits. One important difference is that while an unfavorable (face-on) inclination can ``hide'' the signatures of a planet in RVs, with sufficient precision, planets cannot hide from astrometry regardless of the orbital orientation.  In this sense the detection limits for RVs and HGCA take on slightly different meaning.

Finally, we comment on Gaia parameters that are often useful for identifying close binaries.  The Renormalized Unit Weight Error (RUWE) is a metric used to assess the quality of Gaia's astrometric solutions. Well-behaved single-star solutions typically have values of $\sim$1 (\citealt{Lindegren2021}), with values $>$1.4 being broadly linked to binarity. 51 Peg has a Gaia DR3 RUWE of 1.057, suggesting that there are no close-in stellar companions that are bright enough to distort the PSF and induce systematic error in the astrometry.  
The \texttt{ipd\_frac\_mult\_peak} traces the fraction of Gaia observations with more than one detected peak.  For 51 Peg this value is 0, which further indicates that there are no relatively bright (and previously unknown) comoving stellar companions within the sensitivity limits of Gaia.\footnote{Although \citet{Greaves2006} reported potential common–proper-motion companions to 51 Peg, \citet{Mamajek2010} later showed that these candidates were unrelated background stars.} Finally, we performed a cone search around 51 Peg out to 30\arcsec ($\sim$465 AU) and found no objects exhibiting common proper motion within this radius.\footnote{Note that one faint  source within $30''$ is reported in \textit{Gaia} DR3 (DR3 2835207490908193664). However, although no parallax or proper motion are reported, we find that it is likely a background star; see Section~\ref{sec:WFPC2} for details.}

\subsection{Imaging Analysis}\label{sec:Imaging_Analysis}
\subsubsection{Keck/NIRC2 $L'$ Imaging}\label{sec:NIRC2L}
We obtained high-contrast imaging of 51 Peg on UT 2022 July 12 with the NIRC2 camera at W.M. Keck Observatory. Our observations are taken in $L'$-band ($\lambda_\mathrm{c}$ = 3.776 $\mu$m; $\Delta \lambda$ = 0.7 $\mathrm{\mu m}$) with the Vector Vortex Coronagraph (VVC; \citealt{serabynKeckObservatoryInfrared_2017}) and natural-guide star adaptive optics \citep{wizinowichAstronomicalScienceAdaptive_2013}. Imaging is obtained in pupil-tracking mode to facilitate Angular Differential Imaging (ADI; \citealt{maroisAngularDifferentialImaging_2006}). The Differential Image Motion Monitor (DIMM) measured an average seeing of 0\farcs51 for the night. Imaging consists of sequences of ${\approx}25$ frames obtained with the Quadrant Analysis of Coronagraphic Images for Tip-tilt Sensing (\texttt{QACITS}; \citealt{hubyPostcoronagraphicTiptiltSensing_2015,hubyOnskyPerformanceQacits_2017}) algorithm, which applies tip-tilt corrections to center the star behind the mask. For each sequence, \texttt{QACITS} also obtains short exposures of the unocculted host-star for flux calibration and sky-backgrounds for the science and flux calibration frames. Our science frames each have integration times of $0.18 \, \mathrm{s}$ and 100 coadds. We read out a $512\times512$ pixel subarray which gives a $5\farcs1 \times 5\farcs1$ field of view for our observations. The total integration time and frame rotation for this sequence are $39.0 \, \mathrm{min}$ and $164^\circ$, respectively.

Following flat fielding and dark subtraction, we use the \texttt{L.A. Cosmic} algorithm \citep{vandokkumCosmicRayRejection_2001} to identify and mask cosmic rays. The solution from \citet{serviceNewDistortionSolution_2016} is then applied to correct for geometric optical distortions in the imaging system. The sky-background is modeled and subtracted with seven principal components using the Vortex Image Processing (\texttt{VIP}; \citealt{gomezgonzalezVipVortexImage_2017,christiaensVipPythonPackage_2023}) package. Point-spread function (PSF) subtraction is carried out using the implementation of the Karhunen--Lo\`{e}ve Image Projection \citep[KLIP;][]{soummerDetectionCharacterizationExoplanets_2012} algorithm in \texttt{pyKLIP} \citep{wangPyklipPsfSubtraction_2015}. To tune the PSF subtraction, we carry out injection--recovery across a grid of KL modes (the \texttt{numbasis} parameter); \texttt{movement}, which excludes frames from the PSF model for which a companion would move less than that value in pixels; and the number of \texttt{subsections} on which the PSF subtraction is independently carried out within each annulus. The number of annuli is fixed to four. For each combination of parameters, we inject a source with a separation of $340 \, \mathrm{mas}$, position angle of $63^\circ$, and contrast of $7 \, \mathrm{mag}$, carry out PSF subtraction, and measure the recovered S/N of the source. Additional details on this optimization scheme are presented in Franson et al. (in prep.). The bottom panel of Figure \ref{fig:nirc2_imaging} shows the result of this procedure. We find that the injected source is best recovered with 29 KL modes, 1 subsection, and $\mathtt{movement} = 0\, \mathrm{px}$ and thus adopt this reduction for our analysis.

The top panels of Figure \ref{fig:nirc2_imaging} show our reduced imaging. We do not find any significant (${>}5 \sigma$) sources. Using \texttt{pyKLIP}, we determine a $5\sigma$ contrast curve by measuring the raw contrast level within 2 pixel-wide annuli. Here, we account for small-sample statistics following \citet{mawetFundamentalLimitationsHigh_2014}. The raw contrast level is then corrected for algorithmic throughput through injection--recovery across five azimuthal angles. This produces the contrast curve shown in the middle row of Figure \ref{fig:nirc2_imaging}. We carry out this contrast curve measurement to the edge of the image including separations with partial field of view coverage (2\farcs5--3\farcs4).  Our Keck/NIRC2 imaging places $5\sigma$ upper limits of $\Delta L' = 10.1 \, \mathrm{mag}$ at 0\farcs3, $\Delta L' = 11.4 \, \mathrm{mag}$ at 0\farcs5, $\Delta L' = 12.7 \, \mathrm{mag}$ at $1^{\prime\prime}$, and $\Delta L' = 13.7 \, \mathrm{mag}$ at $2^{\prime\prime}$.  The full contrast curve can be found in Appendix~\ref{sec:contrast_curves}.

\begin{figure}
\hskip -.15 in
{\includegraphics[width=\linewidth]{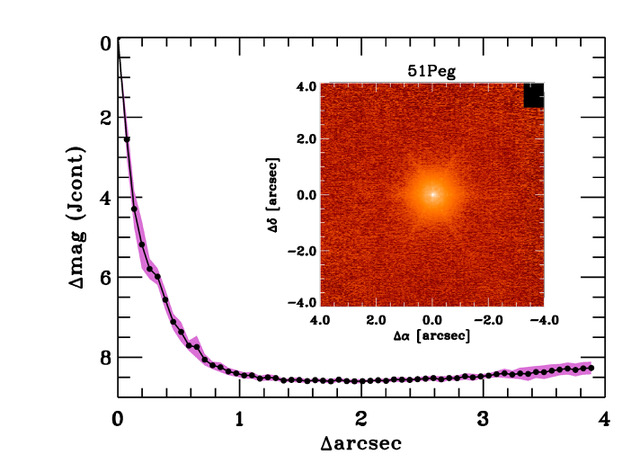}}
\caption{Keck/NIRC2 adaptive optics imaging of 51 Peg in $J_\mathrm{cont}$ (inset). No nearby sources are evident. The 5$\sigma$ contrast is plotted as a function of angular separation. } 
\label{fig:jcont_plot}
\end{figure}

\subsubsection{Keck/NIRC2 J$_\mathrm{cont}$ Imaging}\label{sec:Jcont_imaging}
In addition to deep ADI observations in $L'$ with the VVC, we also targeted 51 Peg with Keck~II/NIRC2 in $J_\mathrm{cont}$ with a wider field of view and no coronagraph.  These NIRC2 observations were obtained on UT 2014 July 22 behind the natural guide star AO system \citep{wizinowich2000} with a 3-point dither pattern that avoids the lower-left quadrant of the detector, which suffers from elevated noise levels. The dither pattern with a $3\arcsec$ step size was performed three times, with each dither pattern offset from the previous dither by $0.5\arcsec$.  NIRC2 was used in the narrow-angle mode with a full field of view of $\sim$10\arcsec $\times$ 10\arcsec and a pixel scale of 9.952~mas pix$^{-1}$ (\citealt{Yelda2010}). \footnote{Note that these data were taken prior to the optical realignment in April 2015.} The Keck observations were made in the narrow-band $J_\mathrm{cont}$ filter $(\lambda_\mathrm{c} = 1.2132~\mu$m; $\Delta\lambda = 0.0198~\mu$m). The integration times for an individual frame was 0.181 sec with 2 coadds per frame, respectively, for a total on-source integration time of 3.258 sec. Flat fields were taken on-sky, dark-subtracted, and median averaged, and sky frames were generated from the median average of the dithered science frames. Each science image was then sky-subtracted and flat-fielded.  The reduced science frames were combined into a single mosaiced image, with a final combined resolution of  0\farcs040 and  0\farcs064, respectively.  

The sensitivity of the final combined AO image were determined by injecting simulated sources azimuthally around the primary target every $20^\circ $ at separations of integer multiples of the central source's FWHM \citep{furlan2017}. The brightness of each injected source was scaled until standard aperture photometry detected it with $5\sigma $ significance.  The final $5\sigma $ limit at each separation was determined from the average of all of the determined limits at that separation and the uncertainty on the limit was set by the rms dispersion of the azimuthal slices at a given radial distance; sensitivities are shown in Figure~\ref{fig:jcont_plot}.  No stellar companions are detected.

\subsubsection{Keck/NIRC2 Wide $H$-band Imaging}\label{sec:Hband_imaging}
We obtained coronagraphic $H$-band imaging of 51 Peg on UT 2026 January 1 at W.M. Keck Observatory with the NIRC2 wide camera and natural-guide star AO \citep{wizinowichAstronomicalScienceAdaptive_2013} to search for stellar and brown dwarf companions at wide separations. The wide camera has a plate scale of $40\, \mathrm{mas}\,\mathrm{pixel}^{-1}$ for a total field of view of $41^{\prime\prime} \times 41^{\prime\prime}$. To avoid saturation, the star is placed behind the $600 \, \mathrm{mas}$-diameter Lyot coronagraph. The DIMM seeing for the night averaged 0\farcs51. We obtained 71 exposures of 100 coadds and $t_\mathrm{int}=0.17 \, \mathrm{s}$ for a total on-source integration time of 20.1 min. Exposures are taken in pupil-tracking mode. The total amount of frame rotation across the sequence is $4.4^\circ$.

Following the same basic data calibration steps as taken for the Keck/NIRC2 VVC imaging (Section \ref{sec:NIRC2L}), we co-register and north-align the frames by fitting the position of the host-star behind the partially transparent coronagraphic mask. For each frame, we estimate and subtract the background through the mask within a circular annulus of 3--5 pixels. The position of the host-star is determined by fitting a two-dimensional elliptical Gaussian. Following an initial measurement, the centroid and sky-background estimate are repeated to produce the final host-star position measurement used to co-register the frames. Aperture photometry of the host star is then carried out within a 2-pixel-radius circular aperture for flux calibration. We then correct for the transmission of the $600\, \mathrm{mas}$ coronagraphic mask, which was empirically calibrated by \citet{bowlerPlanetsLowmassStars_2015} to be $7.51 \pm 0.14 \, \mathrm{mag}$ in $H$. We use observations of the star behind the mask for flux calibration, as the star saturates for the minimum integration time of the NIRC2 wide camera, even for small subarrays. 

\begin{figure}
\begin{center}
{\includegraphics[width=\linewidth]{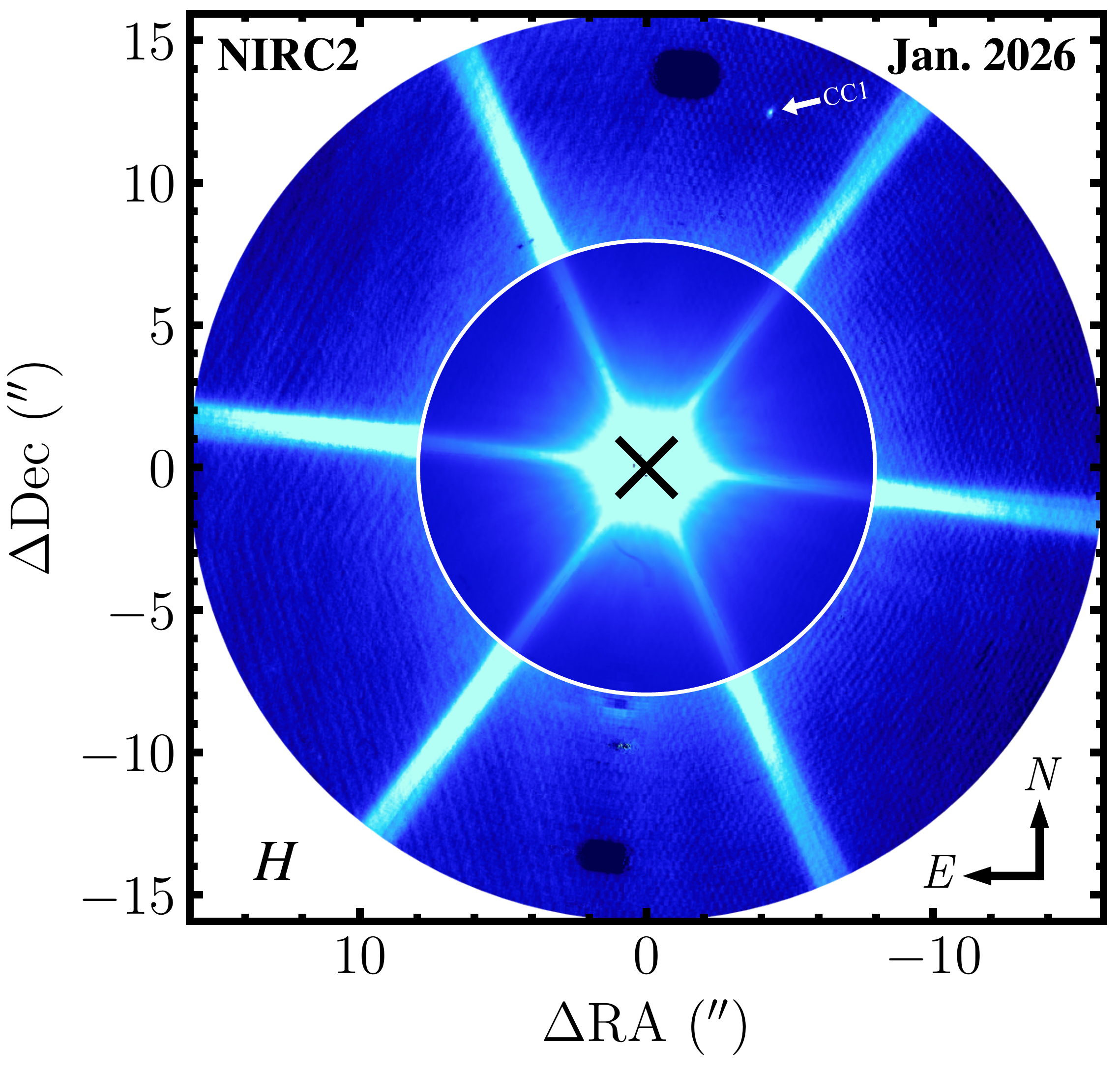}}
\caption{Keck/NIRC2 coronagraphic $H$-band imaging of 51 Peg. These observations are taken with the wide camera for a total field of view of $41^{\prime \prime} \times 41^{\prime\prime}$. The inner $15^{\prime\prime}$ region is shown above. The imaging is shown with an $\mathrm{arcsinh}$ stretch \citep{luptonPreparingRedGreen_2004} and aligned such that north is up and east is left. The dark regions at the north and south of the image are coronagraphic masks. No point sources are identified other than the known background star at ${\approx}14^{\prime\prime}$ (labeled ``CC1'').} 
\label{fig:Hband_imaging}
\end{center}
\end{figure}

Figure \ref{fig:Hband_imaging} shows the median-combined north-aligned reduced image. We recover the background source at ${\approx}14^{\prime\prime}$ separation seen in the HST/WFPC2 imaging and recovered in Gaia (see Section \ref{sec:WFPC2}). No other sources are apparent. A $5\sigma$ contrast curve is measured from the combined image using the standard deviation within 3-pixel-wide concentric annuli. Pixels ${>}5\sigma$ from the median of each annulus are masked for the calculation.

\subsubsection{Hubble Space Telescope/WFPC2 Imaging}\label{sec:WFPC2}
51 Peg was observed by the Hubble Space Telescope's Wide Field and Planetary Camera 2 (HST/WFPC2; \citealt{Holtzman1995}) in three filters---F439W, F555W, and F814W---on UT 1998 May 3 and UT 1998 August 7 as part of a program to detect the source of the RV trend (GO 6845; PI D. Soderblom). The 2-orbit program obtained WFPC2 images of the 51 Peg system at two different orientation angles to search for distant low-mass companions. 
Here, we analyze the reddest of these filters, F814W, as these frames are the deepest and are expected to be most sensitive to cool companions.  
At each orientation, images were taken with integration times of 0.11~s (one frame), 1~s (one frame), 100~s (two frames), and 300~s (two frames), for a total integration time of 811.11s. The two WFPC2 visits were executed at V3 position angles of 21.0$^\circ$ (August) and 81.0$^\circ$ (May), corresponding to a 60$^\circ$ roll difference that positions the diffraction spikes at different locations on the detector. 
51 Peg was centered in the Planetary Camera (PC) CCD, which has a field of view of 36\arcsec $\times$ 36\arcsec and a plate scale of  0\farcs0455 per pixel.
We retrieved both epochs of HST/WFPC2 F814W PC imaging of 51 Peg from the Hubble Legacy Archive.\footnote{Based on observations made with the NASA/ESA Hubble Space Telescope, and obtained from the Hubble Legacy Archive, which is a collaboration between the Space Telescope Science Institute (STScI/NASA), the Space Telescope European Coordinating Facility (ST-ECF/ESA) and the Canadian Astronomy Data Centre (CADC/NRC/CSA).} The science-ready FITS files were generated with the Hubble Legacy Archive Data Release 4 (HLA DR4) and are presented in units of counts per second.\footnote{All the {\it HST} data used in this analysis can be found in MAST: \dataset[10.17909/8rct-8911]{https://doi.org/10.17909/8rct-8911}.}

\begin{figure}
\begin{center}
{\includegraphics[width=\linewidth]{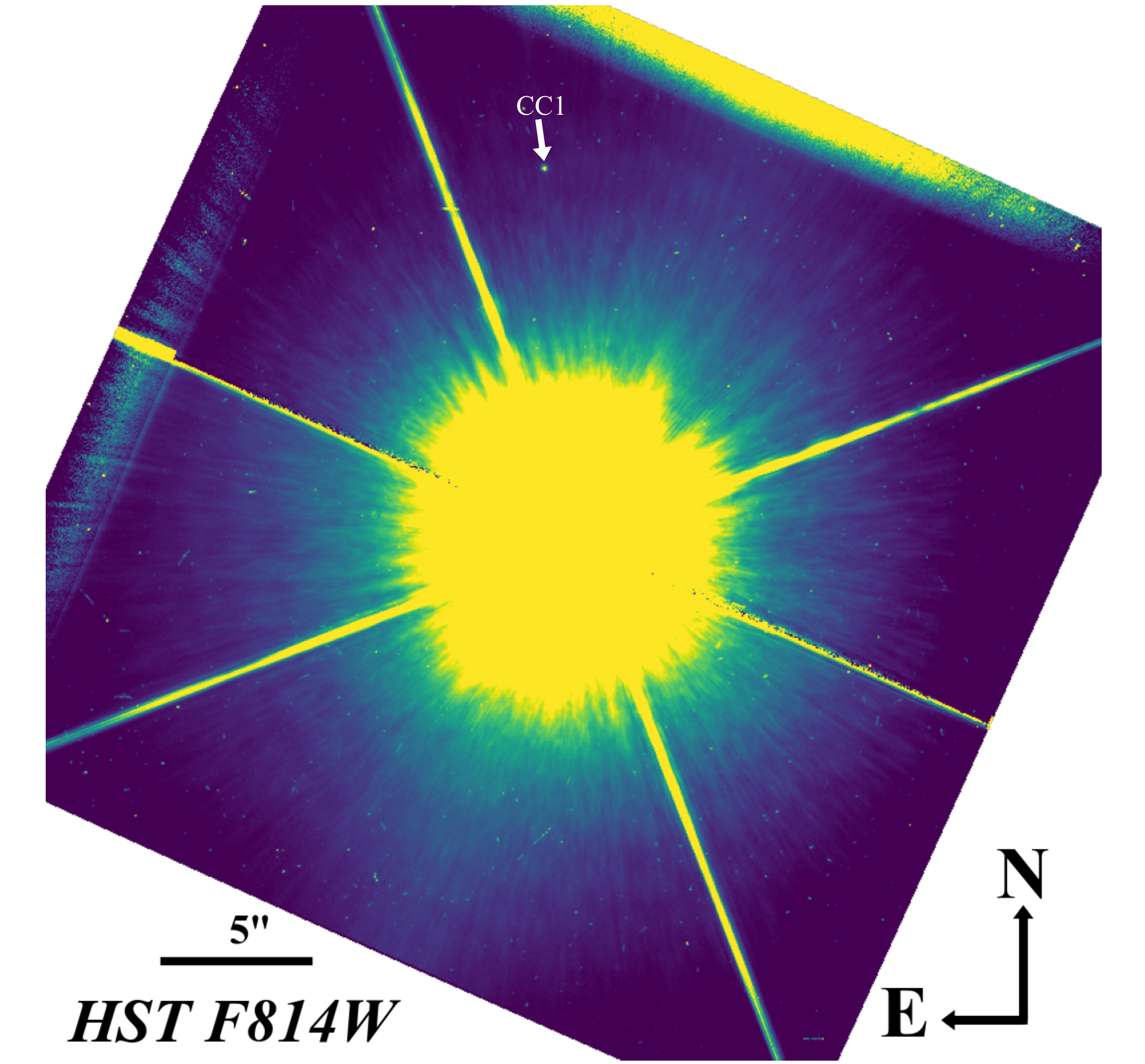}}
\caption{Hubble/WFPC2 imaging of 51 Peg in F814W. The image is oriented such that north is up and east is to the left. The background star ``CC1'' is labeled.}
\label{fig:hubble_814_plot}
\end{center}
\end{figure}

The central core within $\approx$ 1\farcs0 is saturated and the bright PSF wings extend beyond $\approx$ 5\farcs0.  
Figure \ref{fig:hubble_814_plot} shows a faint source $\approx$14'' from the host star at $\mathrm{RA} = 22^\mathrm{h}$ 57$^\mathrm{m}$ 28.08$^\mathrm{s}$ and $\mathrm{Dec} = +20^\circ$ 46$\arcmin$ 22.07$\arcsec$ (ICRS; epoch = J2000) as measured in the August dataset. The closest source within 30" of 51 Peg in Gaia DR3 is Gaia DR3 2835207490908193664 with $\mathrm{RA} = 22^\mathrm{h}$57$^\mathrm{m}$28.05$^\mathrm{s}$ and $\mathrm{Dec} = +20^\circ$46$\arcmin$21.77$\arcsec$ (\citealt{GaiaCollaboration2022}).  No parallax or proper motion measurements are reported but we can readily test if it is comoving with 51 Peg.  51 Peg has moved $\sim$3\farcs9 from 1998.6 (WFPC3 epoch) to 2016.0 (Gaia epoch), while the source has only moved $\sim$ 0\farcs3 during the same baseline (assuming the WFPC2 and DR3 sources are one and the same).  It is clearly a background object. Gaia DR3 2835207490908193664 has a $G_\mathrm{RP}$ magnitude of 18.52 $\pm$ 0.11~mag, which is the closest to the WFPC2 F814W filter. Using aperture photometry, we measure an apparent magnitude of $\sim$19.3 magnitudes in F814W. If this is a main sequence star, it would have an absolute $G$-band magnitude ($M_G$) of $\sim$6~mag and would reside at a distance of $\sim$4.4 kpc (see Figure 2 in \citealt{Gagne2018}).

To generate a contrast curve, we construct a radial noise profile for the August 1998 observations by computing standard deviations within concentric annuli each with 3 pixel widths out to a separation of 13\arcsec. Count rates are converted to flux density using the header value (\texttt{PHOTFLAM} $= 3.522445224719 \times 10^{-19} \ \mathrm{erg \ cm^{-2} \ \AA^{-1} \ e^{-1}}$). We adopt the F814W Vega reference flux  $f_{\lambda,0}^{\mathrm{Vega}} = 1.15119 \times 10^{-9} \ \mathrm{erg \ s^{-1} \ cm^{-2} \ \AA^{-1}}$ from the SVO filter service to convert flux density to apparent magnitude. We report $5\sigma$ contrast limits as a function of angular separation assuming a distance of 15.51 pc (\citealt{BailerJones2021}).

\subsubsection{Previous Imaging of 51 Peg}\label{sec:previous_imaging}

A series of previous investigations with high-resolution imaging have not identified any companions for 51 Peg. \citet{Luhman2002} conducted an adaptive optics imaging survey at Keck to search for companions around RV-detected planet-host stars. Using $H$-band imaging with KCam, they ruled out companions to 51 Peg down to $\Delta H$ $\approx$ 10~mag within 1\arcsec \ and $\Delta H$ $\approx$ 14~mag within 3\arcsec.  \citet{Mason2011} obtained speckle imaging of 51~Peg (HIP~113357) using the US Naval Observatory Speckle Interferometry camera on the 4-m telescope at Kitt Peak National Observatory. They did not detect any sources in their imaging with $\Delta V$ $<$ 3~mag within 0.03-- 1\farcs5 (see their Table~2; \citealt{Mason2011}). \citet{Ngo2017} presented Keck/NIRC2 imaging of 51~Peg (HD~217014), reading out the central 256~pix $\times$ 256~pix (2\farcs5 $\times$ 2\farcs5) region of the array, and detected no stellar companions in their $K_\mathrm{c}$-band observations.
\citet{Roberts2011} conducted an imaging study of companions to planet-hosting stars with the 3.6-m Advanced Electro-Optical System telescope and reported a possible companion $\approx$2\farcs87 from 51 Peg with a $\Delta$$I$ contrast of 10 $\pm$ 0.7 mag. However, only a single epoch was acquired so it is unclear if this potential source is real, and if so whether it shares a common proper motion with the primary star or is instead a background object. Moreover, this candidate has not been recovered in any of the other deeper ground- and space-based imaging. Therefore, we assume that  this source is a spurious signal for the remainder of this analysis.

\begin{figure*}
\begin{center}
\includegraphics[width=\textwidth,height=0.9\textheight,keepaspectratio]{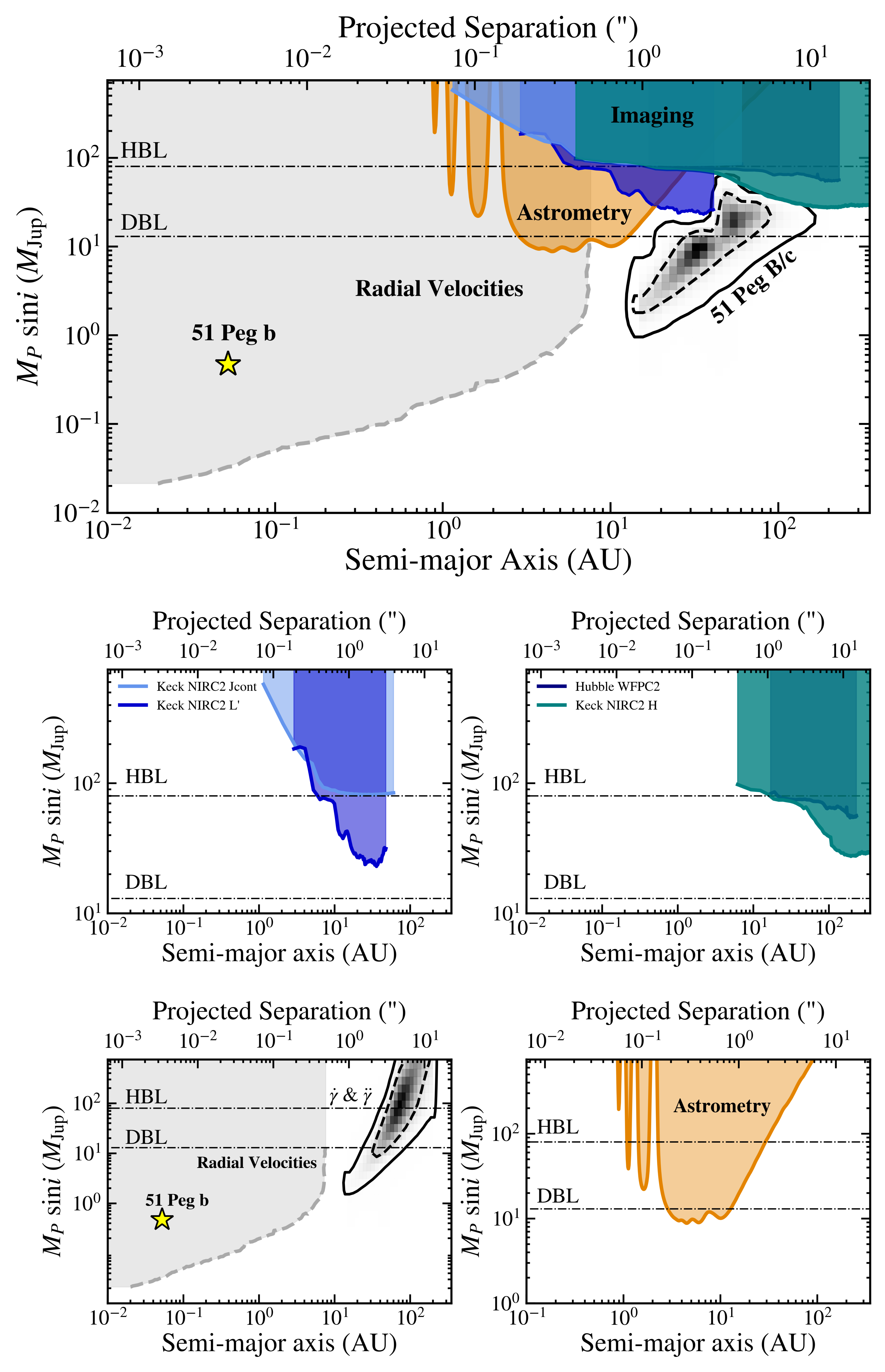}
\caption{Top: Results of the joint sensitivity analysis from a compilation of RVs, astrometry, and high-contrast imaging taken over the past 30 years. 51 Peg b is depicted with the yellow star, and gray contours show constraints on the possible distant outer companion (51 Peg B or c). Most of the mass–separation parameter space is constrained by RVs, so here the $y$-axis is expressed in $m_p \sin i$. However, note that RVs, astrometry, and imaging constraints probe minimum mass, true mass, and model-inferred mass, respectively. Second Row: Masses and separations excluded from the Keck-II/NIRC2 $L'$ and $J_\mathrm{cont}$ AO imaging (left), and both HST/WFPC2 and Keck-II/NIRC2 $H$-band imaging (right). Third Row: The left panel shows the sensitivity to additional companions from RVs.  The dashed gray curve shows the 90$\%$ completeness contour from the RV injection recovery tests.  The contour plot (with 1- and 2-$\sigma$ contour levels) shows regions of parameter space consistent with the reflex motion of the host star based on the measured $\dot{\gamma}$ and $\ddot{\gamma}$ terms. The right panel represents masses and orbits ruled out from the lack of a significant Hipparcos-Gaia acceleration.} 
\label{fig:combined_sensitivty_plot}
\end{center}
\end{figure*}

\section{Results}\label{sec:Results}

\subsection{Synthesizing RVs, Astrometry, and Imaging}\label{sec:Stellar_Activity}
Sensitivity maps from RVs, astrometry, and high-contrast imaging are generated to constrain the allowed and disallowed zones within which potential outer companions can reside. \texttt{RVSearch} (\citealt{Rosenthal2021}) is used to evaluate the detection sensitivity of the long-baseline RV data in minimum mass–separation parameter space. We first subtract out the signal of the inner hot Jupiter, 51 Peg b, as well as the long-term quadratic trend, and then search for periodicity between 1 day ($\sim$ 0.01 AU) and 58,000 days ($\sim$ 30 AU). We adopt a false alarm probability (FAP) threshold of 0.1$\%$ to search for additional periodic signals. The periodogram of the residuals does not result in any peaks that exceed the adopted FAP threshold.\footnote{A separate additional periodogram search can be found in Appendix \ref{sec:periodogram_search}.)} We perform 4,000 injection–recovery tests by injecting synthetic planetary signals into the data and assessing whether they can be successfully recovered. We conclude that a giant planet ($m_p \sin i = 0.3$--13\,$M_\mathrm{Jup}$) would have been detected with high confidence (91\% detection completeness) if present between 0.1 and 10 AU (see Figure \ref{fig:combined_sensitivty_plot}). Saturns and sub-Saturns between 0.1--0.3 $M_\mathrm{Jup}$ can be ruled out to $\approx$0.5 AU. 

Assuming the long-term RV signal reflects dynamical reflex motion from a distant body, its amplitude and shape provide information about pairings of mass and period of the distant body.
Here we explore the possible range of phase space in which the outer companion may reside.
In general, a long-term linear RV trend corresponds to a power law in (minimum) mass and semimajor axis (e.g., \citealt{Torres1999}); companions with longer periods must be more massive to induce the same signal.
With quadratic curvature, the behavior of this mass-separation degeneracy is broader and more structured.  For low companion masses, the resulting constraint shows signs of both the $m$ $\propto$ $a^2$ scaling (from $\dot{\gamma}$) and $m$ $\propto$ $a^{7/2}$ (from $\ddot{\gamma}$; see Appendix~B of \citealt{VanZandt2024} for details). In principle, this behavior continues to high masses and wide separations, but direct imaging can complement these constraints by ruling out distant stellar and substellar companions.  Similarly, short orbital periods can be excluded by considering the time baseline of the RVs, as quadratic behavior implies an orbital period of at least a few times the duration of the observations.

We use \texttt{ethraid} (\citealt{VanZandt2024}) to assess how the $\dot{\gamma}$ and $\ddot{\gamma}$ radial acceleration terms map to constraints in companion mass and separation. Orbits are each sampled from a log-uniform prior on the semi-major axis from $10^{1}$--$10^{2.5}$ AU, a log-uniform prior on the companion mass from $10^{-1}$--$10^{3}$~{$M_\mathrm{Jup}$}, a Uniform prior on eccentricity from 0--1, a Uniform prior on the argument of periastron\footnote{Note that the argument of periastron is sampled for the outer companion not the host star.} from 0--2$\pi$,  a Uniform prior on the the mean anomaly at a reference epoch from 0--2$\pi$,  and the inclination of the orbit sampled in $\cos i$ with a Uniform distribution from 0--1. Prior distributions are then weighted by Equation 11 in (\citealt{VanZandt2024}).

As seen in Appendix \ref{sec:Additional_Statistical_Tests}, the $\dot{\gamma}$ and $\ddot{\gamma}$ terms show strong covariance. This is expected, as the coefficients of the first and second terms describing a parabola will vary together in a similar way as the slope and $y$-intercept of a line are correlated. To properly account for covariance between the two terms, we modify the likelihood function in \citet{VanZandt2024} to a more general bivariate Gaussian. The resulting updated log-likelihood function, $\ln\mathcal{L}_{\mathrm{RV}}(\theta)$, is given by

\begin{equation}\label{eqn:covariance_liklihood_equation}
\begin{aligned}
 \ln\mathcal{L}_{\mathrm{RV}}(\theta) = 
-\frac{1}{2(1-\rho^2)} \\
\Bigg(
\frac{(\dot\gamma_{\rm data}-\dot\gamma)^2}{\sigma_{\dot\gamma}^2}
-2\rho\,\frac{(\dot\gamma_{\rm data}-\dot\gamma)(\ddot\gamma_{\rm data}-\ddot\gamma)}{\sigma_{\dot\gamma}\sigma_{\ddot\gamma}} \\
+\frac{(\ddot\gamma_{\rm data}-\ddot\gamma)^2}{\sigma_{\ddot\gamma}^2}
\Bigg),
\end{aligned}
\end{equation}

\noindent where $\theta$ denotes the set of samples drawn from an importance function defined over the orbital parameters, $\dot{\gamma}$ and $\ddot{\gamma}$ are the coefficients of the linear and quadratic acceleration terms, and $\sigma_{\dot\gamma}^2$ and $\sigma_{\ddot\gamma}^2$ are their respective errors. The correlation coefficient, $\rho$, is defined as:
\begin{equation}
\rho \equiv 
\frac{\gamma_{\rm cov}}
{\sigma_{\dot{\gamma}}\,\sigma_{\ddot{\gamma}}}.
\end{equation}

\noindent
Here, $\gamma_{\rm cov}$ is the covariance between the two terms.  
For our measurements, we find $\gamma_{\rm cov}$ = $-4.39\times10^{-14}$ and $\rho$ = --0.99, indicating a strong negative correlation.

Using the open-source package \texttt{species} \citep{Stolker2020}, along with the ``BT-Settl'' \citep{Allard2014} and ``ATMO'' \citep{Phillips2020} evolutionary models, we convert the NIRC2 and Hubble Space Telescope contrast and sensitivity curves (see Section \ref{sec:Potential White Dwarf Companions} and Appendix~\ref{sec:contrast_curves}) to ultracool companion mass limits assuming an age of 6~Gyr (e.g., \citealt{Takeda2007}; \citealt{Isaacson2024}; see Section~\ref{sec:Stellar_Activity}). 
The WFPC2 sensitivity curve is derived in units of $\mathrm{erg\ s^{-1}\ cm^{-2}\ \AA^{-1}}$ as a function of angular separation.  This is converted to apparent magnitudes using the appropriate Vega zero point for the WFPC2 F814W filter following the WFPC2 documentation available through STScI.  To convert the apparent magnitude to a contrast with respect to the host star for the \texttt{species} package, we use the solar spectrum from \citet{Rieke2008} as a proxy for 51 Peg (which shares the same G2 spectral type).  Scaling the solar spectrum to the distance of 51 Peg at 15.51~pc and computing synthetic photometry through the F814W filter profile yields an apparent magnitude of F814W = 5.05~mag for 51 Peg.  From this we compute a contrast curve in $\Delta$F814W and translate this into constraints on companion mass.

\begin{figure*}
\begin{center}
{\includegraphics[height=0.22\textheight, keepaspectratio]{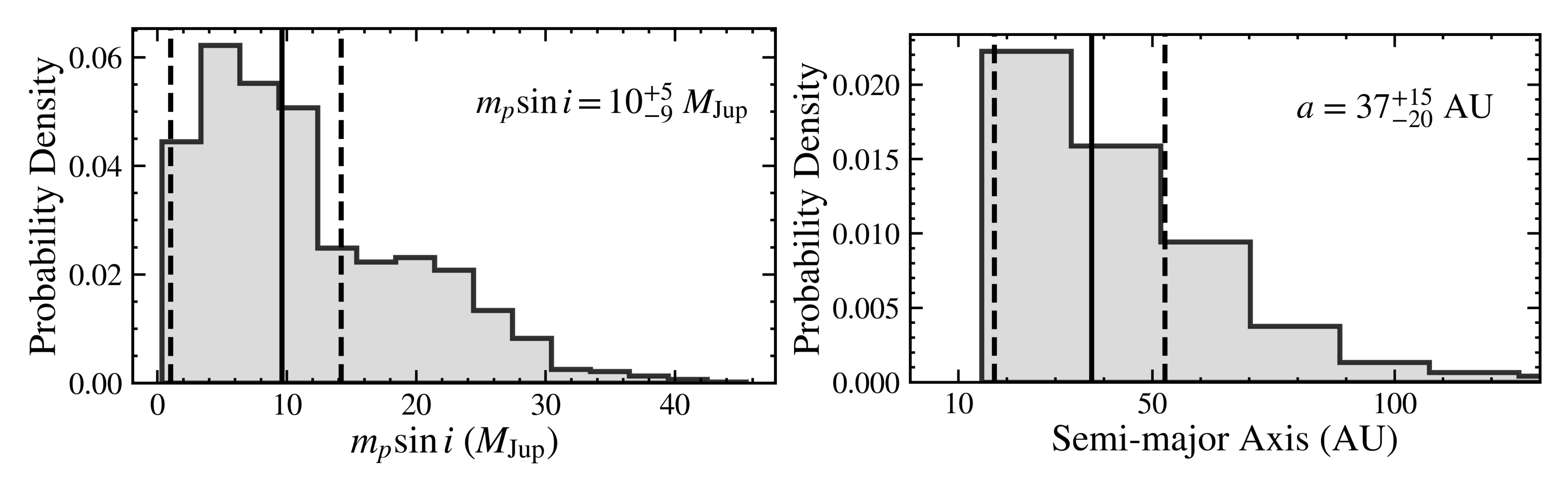}}
\vskip -.1 in
\caption{Marginalized distributions of companion mass and semi-major axis from the final joint constraints. The inferred minimum mass spans $\approx$1--50~$M_\mathrm{Jup}$ and the semi-major axis spans $\approx$15--170 AU. The median value is reported along with the upper and lower bounds of the 68.3$\%$ highest density interval regions. }
\label{fig:marginilized_mass_sep_histograms}
\end{center}
\end{figure*}

The deep NIRC2 $L'$ high-contrast imaging allows us to probe down to brown dwarf mass companions out to 40 AU ($\approx$$3^{\prime\prime}$). The shallower NIRC2 $J_\mathrm{cont}$ imaging probes stellar-mass companions out to 60 AU ($\approx$$4^{\prime\prime}$). The HST WFPC2 and NIRC2 $H$-band imaging are sensitive to stellar-mass companions and high-mass brown dwarfs at separations up to 230 AU ($\approx$$15^{\prime\prime}$) and 390~AU ($\approx 25^{\prime\prime}$), respectively. Altogether, high-contrast imaging of 51 Peg rules out stellar companions between projected separations of $\sim$5 to 390 AU and intermediate-mass brown dwarfs from $\sim$10 to 40 AU. This phase space intersects with constraints on the outer companion's mass and separation from RVs, effectively truncating the high-mass (stellar) possibility and tightening posterior constraints on its mass and orbit.

The lack of an astrometric detection from \textit{Hipparcos}-\textit{Gaia} excludes close-in brown dwarfs and high-mass planets from $\sim$2--20 AU ($\sim$0.1--1$^{\prime\prime}$) and a broader range of stellar companions between $\sim$2--60 AU ($\sim$0.1--4$^{\prime\prime}$).  This does not formally intersect with the predicted masses and separations of the outer companion from RV constraints, but in general astrometric upper limits offer unique information that is complementary to upper limits from RVs.  Companions will be inaccessible to RVs if their orbital inclinations are low and happen to lie near the plane of the sky, but astrometry is sensitive to planets with any orientation.  We can therefore be quite confident that this region is bare of companions with these masses and orbits.

After taking into account these upper limits from other methods, the resulting posterior distribution of allowed masses and separations for the companion are $m_p \sin i$ = 10$^{+5}_{-9}$ $M_\mathrm{Jup}$ and $a$ = 37$^{+15}_{-20}$ AU (Figures \ref{fig:combined_sensitivty_plot}- \ref{fig:marginilized_mass_sep_histograms}). The combination of these separate yet complementary datasets excludes a substantial portion of the parameter space in the 51 Peg system, suggesting that it is largely devoid of additional massive companions. If the observed radial acceleration is due to Doppler reflex motion rather than instrumental drift (see Section~\ref{sec:hamilton_trend_artifact}), the combined constraints from these datasets suggest that the distant companion is a high-mass Jovian planet, ``51 Peg c,'' or a brown dwarf, ``51 Peg B.''

\section{Discussion}\label{sec:Discussion}

\subsection{On the Hamilton Trend: Instrument Drift?}\label{sec:hamilton_trend_artifact}
Shallow RV trends can be difficult to distinguish from instrumental drifts or systematic effects, which can be caused by subtle changes in the spectrograph’s thermal and mechanical state, evolving illumination of the entrance fiber, detector-related systematics, and imperfections in wavelength calibration that introduce time-dependent shifts in the measured line centroids (e.g., \citealt{Fischer:2016hl}; \citealt{Burt:2025aa}). 
To test the nature of the Hamilton Doppler trend, which drives the overall curvature in the best-fitting model, we analyze a set of RV standard stars monitored with both the Hamilton and HIRES spectrographs using the measurements reported in \citet{Fischer2014} and \citet{Rosenthal2021}, respectively.

Precision RV standards are stars with stable (usually flat) RVs at the m s$^{-1}$ level and well-characterized long-term behavior. They serve as references to track instrumental drifts over time and to define the zero-point of spectrographs. The RV standard stars analyzed here are first identified in \citet{Chubak2012} and cross-matched with \citet{Fischer2014} and \citet{Rosenthal2021}. We then restrict the sample to stars with at least five Hamilton RV measurements, observations obtained after 1995-01-01 (UTC) to match the baseline of the 51 Peg RVs, a minimum observational baseline of 8 years, and no known RV trends in the corresponding HIRES RVs. Altogether, this results in 19 stars that can be used to assess whether the Hamilton RVs suffer from systematic trends compared to HIRES, which has a long history of precise and stable RVs since the detector upgrade in 2004 (\citealt{Butler:2017km}). Details of the individual stars and a comparison of slopes can be found in Appendix~\ref{sec:RV_Standard_Analysis}.

We find that the Hamilton RVs of the standard stars show drifts comparable to the long-term trend in the 51 Peg data.  The mean slope is --0.53 m s$^{-1}$ yr$^{-1}$ with a standard deviation of 0.90  m s$^{-1}$ yr$^{-1}$. For comparison, we analyze the same RV standard stars using HIRES data from \citet{Rosenthal2021} and find that the width of the instrumental drift distribution is much narrower than that of the Hamilton data (see Appendix \ref{sec:RV_Standard_Analysis}): the mean slope is --0.05 m s$^{-1}$ yr$^{-1}$ with a standard deviation of 0.32  m s$^{-1}$ yr$^{-1}$. The Hamilton time series appear to experience instrument-related systematic slopes up to about $\pm$2 m s$^{-1}$ yr$^{-1}$ compared to the HIRES datasets.

The Doppler trend observed in the 16-year Hamilton dataset for 51 Peg (${\dot{\gamma}}$ = --1.35$^{+0.01}_{-0.01}$ m $\text{s}^{-1}$ yr$^{-1}$) is comparable in scale to the drifts measured in the RV standard stars, making it challenging to distinguish whether the observed trend for 51 Peg originates from a long-period companion or from potential instrumental systematics. If the observed radial acceleration is due to instrumental drift, the analysis in this study places strong constraints on what is not present in this system.  In this scenario, 51 Peg b has no outer Jovian companions out to about 10 AU and most brown dwarf masses are excluded out to several tens of AU. The absence of additional companions may favor a disk-driven migration scenario for the inner hot Jupiter, 51 Peg b.

So far, we have explored false positive scenarios for the Hamilton spectrograph. It is also instructive to examine true positive cases in which shallow RV trends, comparable in scale to the Hamilton drifts measured in the RV standard stars, revealed low-mass substellar companions. For instance, GL 758 (HD 182488) is a nearby G8 star hosting a T7–T8 brown dwarf companion Gl 758 B (\citealt{Thalmann2009}; \citealt{Vigan2016}; \citealt{Bowler2018}; \citealt{Brandt2019b}). The $\approx$38~$M_\mathrm{Jup}$ substellar companion induces a linear RV trend in the host star of $-2.25 \pm 0.09$~m~s$^{-1}$~yr$^{-1}$ based on 118 RVs obtained with the Tull/2.7~m, $-2.82 \pm 0.03$~m~s$^{-1}$~yr$^{-1}$ from 262 RVs obtained with Keck/HIRES, and $-4.24 \pm 0.07$~m~s$^{-1}$~yr$^{-1}$ from 250 RVs obtained with APF/Levy \citep{Bowler2018}. (Note that each of these datasets samples a later midpoint epoch, and the RV trend is becoming steeper over time.) Our analysis of 36 Hamilton RVs \citep{Fischer2014} reveals a slope of $-1.88 \pm 0.57$~m~s$^{-1}$~yr$^{-1}$, which is most consistent with results from the Tull spectrograph and overlaps in time with those observations. This test case provides evidence supporting the interpretation that the trend observed in the 51~Peg system is consistent with stellar reflex motion.

Where does this leave us?  The Hamilton RVs show clear signs of long-term drifts, but conversely, independently measured long-term RV trends have been recovered at the few-m s$^{-1}$ level.  We conclude that the shallow acceleration from 51 Peg seen in this early Hamilton dataset---which has been noted by many previous studies in the past---is consistent with a systematic imprint from the spectrograph or its associated RV calibrations, but it remains plausible that the signal is real.  Below, we further explore the implications of the latter scenario.

\subsection{Reflex Motion or an Activity-related RV Signal?}\label{sec:Stellar_Activity}

Active regions suppress a star’s convective blueshift, resulting in a net redshift in RVs (\citealt{Burt2025}). This can produce a correlation between long-term RVs and $\log R'_{\mathrm{HK}}$ values (\citealt{Isaacson2010}; \citealt{Lovis2011};  \citealt{Robertson2013}; \citealt{GomesdaSilva2021}; \citealt{Rosenthal2021}), which trace chromospheric activity levels. Precise RVs are therefore sensitive to long-term stellar magnetic activity cycles, which may induce slowly changing RV variations with amplitudes and timescales comparable to those of Jupiter-like planets orbiting at several AU (\citealt{Lovis2011}; \citealt{Robertson2013}). In general, it is difficult to disentangle the signal of a long-period planet and an activity cycle, especially if RVs and activity indicators only span a fraction of a period cycle (\citealt{Isaacson2024}). 
This raises the question of whether the slight curvature evident in the RVs of 51 Peg represent gravitational reflex motion from a distant body or are instead the result of a long-term stellar activity cycle.

Sun-like stars undergo magnetic braking over long timescales which slows their rotation periods, reduces their magnetic dynamos, and results in a decay in the associated tracers of activity---H$\alpha$ line strength, X-ray and UV luminosity, \ion{Ca}{2} H and K line strenghts, starspot surface covering fractions, and photometric modulations (\citealt{Babcock1961}; \citealt{Hall2008}; \citealt{MamajekHillenbrand2008}; \citealt{vanSaders2016}; \citealt{Charbonneau2020}; \citealt{Metcalfe2024}).  At 4.6~Gyr, the Sun has an 11-year activity cycle and a corresponding 22-year magnetic cycle (\citealt{Hathaway2015}; \citealt{Jeffers2023}). The Sun also experiences less dominant, even longer-term cyclical behavior such as the 80-year Gleissberg cycle (\citealt{Gleissberg1939}).\footnote{\citet{SaarBrandenburg1999} suggest that long-term activity trends in stars may represent segments of long-period (50--100 yr) Gleissberg-like cycles that have not yet been resolved with current observational baselines.} There are signs that these trends continue at even older ages: \citet{Lovis2011} analyzed the magnetic activity cycles of 304 FGK stars for $\approx$7 years using the HARPS spectrograph (\citealt{Mayor2003}) and found that stars with a mean $\log R'_{\mathrm{HK}}$ near $-5.0$ rarely exhibit magnetic cycles as strong as the Sun. This suggests that magnetic activity continues to diminish with stellar age and stars lose detectable cycles once they reach extremely low activity levels (e.g., \citealt{Isaacson2025}). 

51 Peg has been the focus of nearly 50 years of chromospheric activity monitoring at Mount Wilson Observatory (\citealt{Wilson1968}; \citealt{Metcalfe2024}). It is an inactive main-sequence G2 star with measured $\log R'_{\mathrm{HK}}$ values ranging from $-5.090$ to $-4.989$ (\citealt{Henry2000}; \citealt{BoroSaikia2018}; \citealt{Baum2022}; \citealt{Isaacson2024})---even less active than the Sun ($\log R'_{\mathrm{HK}}$ = --4.984; \citealt{Egeland2017}).  As a result of its unusually low activity, 51 Peg has been proposed as a Maunder minimum candidate:\footnote{Maunder Minimum refers to the period around 1645 to 1715 during which sunspots became exceedingly rare and where solar activity levels were likely low (see \citealt{Hathaway2015}).} \citet{Poppenhager2009} analyzed \texttt{XMM-Newton} and ROSAT X-ray measurements and found that 51 Peg has consistently low coronal emission, similar to the flat, unspectacular \ion{Ca}{2} H and K line emission measured over many decades. 

A detailed analysis of the individual spectra associated with each RV measurement is beyond the scope of this paper.  However, we have compared the RVs and stellar $S$-index values---which map the strength of \ion{Ca}{2} H and K emission lines (\citealt{Isaacson2010})---as a function of time. We analyze the $S$-value time series spanning 6, 12, and 17 years for the Hamilton (\citealt{Isaacson2010}), APF, and HIRES (\citealt{Rosenthal2021}) datasets, respectively, by fitting a linear trend in time and compare it to a constant (flat) model using the BIC (see Appendix \ref{fig:S_value_time_series}). We find that a linear trend is strongly favored in the APF dataset ($\Delta$BIC = 23.1), while the HIRES ($\Delta$BIC = 6.4) and Hamilton ($\Delta$BIC = 1.7) datasets favor a flat model. We also find correlation coefficients between the $S$-index and residual RV measurements (after subtracting the signal from 51 Peg b) of 0.04, --0.03, and 0.01 for the Hamilton, APF, and HIRES datasets, respectively, and similarly compare linear and flat models using the BIC. In all three datasets, the flat model is statistically preferred, indicating no significant correlation between stellar activity and the observed long-term RV variations. In summary, with the exception of a trend in the APF $S$-value time series, we find no significant correlation between stellar activity and time or with the RV measurements. 

However, because the amplitude of the trend is so small, it remains possible that activity could nevertheless explain the observed signal but simply be too weak to induce noticeable imprints in the $S$ index. If this is the case, we can estimate whether the inferred activity cycle (based on the star's rotation period) is consistent with the implied period of the RV signal.  This requires constraints on several fundamental properties of 51 Peg---most importantly its age and, to the extent possible, its rotation period.

The precise age of 51 Peg has been challenging to  determine but isochrones and activity indicators consistently indicate it is older than the Sun. For instance, \citet{Takeda2007} estimated an age of 6.8$^{+1.6}_{-1.5}$ Gyr using  precise stellar parameters from high-resolution spectroscopy and stellar evolution models, and \citet{Isaacson2024} found an age of 6.65 Gyr from $\log R'_\mathrm{HK}$-age relations. \citet{Holmberg2009}, \citet{Brewer2016}, and  \citet{GaiaCollaboration2022} estimate ages of $6.0^{+2.1}_{-2.0}$~Gyr, $4.7^{+1.0}_{-1.1}$~Gyr, and  $7.8^{+1.1}_{-1.0}$~Gyr, respectively, based on stellar evolutionary tracks. 

We calculate similar ages with a long-term average of activity indicators and gyrochronology relations. Based on 8 published measurements of $\log R'_\mathrm{HK}$ spanning 30 (1993--2023) years,\footnote{We combine published $\log R'_{\mathrm{HK}}$ measurements from \citet[$-5.068$ dex]{Henry2000}, \citet[$-5.071$ dex]{Gray2003}, \citet[$-5.080$ dex]{MamajekHillenbrand2008}, \citet[$-5.054$ dex]{Isaacson2010}, \citet[$-5.060$ dex]{Brewer2016}, 
\citet[$-4.989$ dex---which is an average of five measurements]{BoroSaikia2018}, \citet[$-5.090$ dex]{Radick2018}, and \citet[$-5.050$ dex]{Isaacson2024}.} we find an average activity level of $\log R'_\mathrm{HK} = -5.058 \pm 0.03~\mathrm{dex}$. 
This corresponds to an age of 7.6 $\pm$ 1.5 Gyr using Equation 3 from \citet{Mamajek2010}, which includes an rms of 0.07 dex around their best-fit relation. For this study we have adopted a characteristic age of 6~Gyr, with typical values in the literature spanning $\sim$5--9~Gyr.

\citet{Noyes1984} surveyed slowly rotating main-sequence stars and found a connection between the stellar rotation period ($P_{\mathrm{rot}}$) and the activity cycle period ($P_\mathrm{cyc}$; see also \citealt{SaarBrandenburg1999}; \citealt{BohmVitense2007}; \citealt{Metcalfe2016}; \citealt{BoroSaikia2018}).  
\citet{Olah2016} analyzed the relationship between stellar rotation periods, effective temperatures, chromospheric activity, and stellar age for a larger sample of G and K dwarfs. They found that for old ($\gtrsim$2~Gyr) solar analogs, rotation periods and activity cycles scale as $P_{\mathrm{cyc}}$ $\sim$ $10^{2} \times P_{\mathrm{rot}}$.  That is, both cycles appear to decay together in such a way that their ratio is approximately constant over time.

The rotation period of 51 Peg is not well established, so the duration of the expected activity cycle for this star is unclear.  There have been several reported rotation periods of 21 days, 22.6 days, and 37 days derived from Ca\,\textsc{ii} H\&K and photometric measurements (\citealt{Baliunas1996}; \citealt{Henry2000}; \citealt{Simpson2010}).  However, no single period has been reliably recovered and the amplitudes of these signals are small relative to the noise levels of these time series datasets.
Using the empirical relation for Sun-like stars from \citet[their Equation 1]{Angus2015}, 

\begin{equation}\label{eqn:gyrochronology_equation}
P = A^{n} \times a (B - V - c)^{b},
\end{equation}

\noindent we can estimate the stellar rotation period, $P$ (in days), where $A$ is the stellar age (in Myr), $B$ and $V$ are the $B$- and $V$-band magnitudes, and $a$, $b$, $c$, and $n$ are dimensionless parameters calibrated against star clusters and well-studied asteroseismic benchmarks. Given the $B - V$ color of 0.67~mag for 51 Peg (\citealt{Isaacson2010}) and an estimated age of 6~Gyr, the implied rotation period is $\sim$30~days and $\sim$37~days for ages of 6 and 9~Gyr, respectively.

How does this compare with other mature Sun-like stars?
\citet{Santos2021} measured the rotation periods of $\approx$19,000 main-sequence G stars in the Kepler long-cadence data and found that nearly all stars with detectable periodic modulations ($\approx$30\% of the full sample) have $P_{\mathrm{rot}}$ $<$ 50 d. Only $\approx$50 G dwarfs---or about 0.3\% of the subsample---have $P_{\mathrm{rot}}$ $>$ 50 d. Similarly, only $\sim$9\% of the sample have periods longer than 30 d.
Long rotation periods are therefore rare, although we note that light curve analyses tend to underrepresent extremely slow rotating stars due to biases in spot-induced variability amplitudes. 

In a recent study, \citet{Gaidos2025} analyzed \textit{TESS} two-minute cadence photometry of 51 Peg in Sectors 56 and 83 and found a low-amplitude 4.55-day periodic signal at the 50 ppm level distinct from the orbital period of the planet. They conclude that the signal could be explained as a planet-induced forced oscillation if the stellar rotation period were 60 days; however, they also note that such a slow rotation is atypical among similar G dwarfs with measured rotation periods.

Despite the lack of a robust rotation period, we can estimate what the activity cycle would be and compare this with the long-term RV signal \emph{if} the true rotation period of 51 Peg is comparable to these reported values.
For a rotation period of 22~days, the corresponding activity cycle would be approximately 6~years (assuming $P_\mathrm{cyc}$/$P_\mathrm{rot}$ $\sim$ 100). Longer rotation periods of 30, 40, and 50~days yields activity cycles of roughly 8, 11, and 14~years, respectively. All of these are substantially shorter than the baseline of the RV observations (31 years) and the implied period from the slowly changing RVs (at least several times this baseline, $\gtrsim$60 years).
Approaching this from the other direction, if we attribute this lower limit of $\sim$60~years to an activity cycle, the corresponding rotation period would be $\gtrsim$219~days.  If we rearrange Equation~\ref{eqn:gyrochronology_equation} and substitute in this rotation period, this yields an age of over 200~Gyr---far greater than the Hubble time.  
We conclude that the RV signal, if real, is most readily explained by a distant body and is irreconcilable with an activity cycle based on current empirical constraints for old solar analogs.

\subsection{Companions to Hot Jupiters}\label{sec:Hot Jupiter Companions}

Recent studies have shown that hot Jupiters are frequently accompanied by outer companions, suggesting that a significant fraction of them may have been dynamically excited to highly eccentric orbits early in their lifetimes, later evolving into close-in planets through tidal circularization (\citealt{rasioford1996}; \citealt{chatterjee2008}). Long-term RV surveys have found that $\approx$70$\%$ $\pm$ 8$\%$ of stars harboring hot Jupiters also host planetary companions with masses between 1--13 $M_\mathrm{Jup}$ and orbital semi-major axes between 1--20 AU (\citealt{Knutson2014}; \citealt{Bryan2016}). In addition, AO imaging surveys have found that 47$\%$ $\pm$ 7$\%$ of hot Jupiter host stars also harbor stellar companions with semimajor axes between 50 and 2000 AU (\citealt{Ngo2016}). \citet{Zink2023} found evidence that in planetary systems containing both a hot Jupiter and an outer giant companion, co-planar high-eccentricity migration is most likely the dominant mechanism influencing the orbit of the inner giant planet. They also found that the companions are almost always $>$3 times the mass of the hot Jupiters.

The 51 Peg system is consistent with these results whereby a more massive companion may have dynamically scattered the (proto-)hot Jupiter 51 Peg b--initially at a much wider separation---into a highly eccentric orbit, which then shrunk and circularized through tidal interactions with the host star. The observed trend that metal-rich host stars tend to harbor more giant planets (\citealt{Santos2004}; \citealt{FischerValenti2005}; \citealt{Johnson2010}; \citealt{Petigura2018})---potentially increasing the chances of dynamical scattering events (\citealt{DawsonMurray-Clay2013}; \citealt{Morgan2025a})---may also apply to this well-studied system, which itself is metal-rich ([Fe/H] = +0.21 dex; \citealt{Rosenthal2021}). 

\citet{Birkby2017} derived a constraint on the orbital inclination of  51 Peg b (70$^\circ$ $<$ $i_{p}$ $<$ 82.2$^\circ$) from the planet's orbital velocity and induced semi-amplitude. This is consistent with an estimate on the the lower limit on the projected stellar inclination of the host star of 70$^{+11}_{-30}$$^\circ$ (\citealt{Simpson2010}) suggesting spin orbit alignment. If the outer companion is coplanar with the inclination of 51 Peg b, this would suggest an origin related to coplanar planet-planet scattering (\citealt{Petrovich2015}; \citealt{Zink2023}). However, if the outer companion is highly inclined relative to the orbit of 51~Peg~b, this may indicate the presence of ZLK oscillations (\citealt{Wu2007}; \citealt{Naoz2012}). The relative alignment of 51 Peg B/c with the stellar host and hot Jupiter is currently unconstrained.  Upcoming opportunities with Gaia DR4 and DR5, and high-contrast imaging (for instance, with JWST or the next generation of ground-based Extremely Large Telescopes), offer an opportunity to recover the companion and establish its orbital inclination.  This would provide additional insight into the dynamical history of this system and help further assess the influence of a higher-mass companion on the migration of 51 Peg b.

\subsection{Potential White Dwarf Companions}\label{sec:Potential White Dwarf Companions}

Ninety-eight percent of all stars ultimately evolve into white dwarfs with a typical mass of $\approx$0.6 ~$M_{\odot}$ (\citealt{Liebert2005}; \citealt{Kepler2007}) and a range spanning $\sim$0.2--1.4~$M_{\odot}$  (\citealt{Weidemann1983}; \citealt{Winget2008}; \citealt{Pelisoli2025}).
In binary systems, the more massive component will evolve more rapidly, leaving behind a white dwarf for progenitor masses under $\approx$8~$M_{\odot}$.
The class of ``Sirius-like binaries'' with a spatially resolved white dwarf companion to a main-sequence star are especially useful as tests of white dwarf cooling models if their dynamical masses can be determined from the reflex motion on the host star (e.g., \citealt{Brandt2019}; \citealt{Bowler2021B}). So far, in our analysis of the imaging constraints we have assumed the companion is a main-sequence star, brown dwarf, or giant planet; here we assess whether the long-term trend could instead be consistent with a distant white dwarf.

 Our Keck/NIRC2 $H$-band contrast curve is used to compare the sensitivity of the observations to the expected absolute magnitude of a typical white dwarf companion. Assuming a white dwarf mass of 0.6~$M_{\odot}$, the initial-to-final mass relation from \citet{Cummings2018} implies a progenitor mass of $\sim$1.5~$M_{\odot}$ and a progenitor main-sequence lifetime of $\sim$3~Gyr (\citealt{Paxton2011}; \citealt{Choi2016}; \citealt{Dotter2016}).  This would imply a white dwarf cooling age of $\sim$3~Gyr given the adopted age of $\sim$6~Gyr for 51 Peg.
  Based on MESA Isochrones and Stellar Tracks (MIST) white dwarf cooling models (\citealt{Tremblay2011}; \citealt{Bauer2025}), white dwarfs with characteristic cooling ages of 3~Gyr and masses of $\sim$0.5--0.8~$M_{\odot}$ are expected to have absolute $H$-band magnitudes of $\sim$13.1--13.5~mag. Our Keck $H$-band imaging reaches absolute magnitudes of $\approx$15~mag at 3\arcsec, 17~mag at 5\arcsec, and 19~mag at 10\arcsec. The contrast curve extends to $\approx$25\arcsec---well beyond the predicted mass and separation of the potential companion implied from the RVs. Therefore, it is  unlikely that a bound white dwarf companion is responsible for the observed shallow acceleration.

\section{Conclusion}\label{sec:Conclusion}
We have presented evidence for a potential high-mass planet or brown dwarf outer companion in the 51 Peg system by synthesizing
new and previously published RVs, ground- and space-based high-contrast imaging, and absolute astrometry. Below is a summary of our main results.

\begin{itemize}[topsep=0pt, partopsep=0pt, itemsep=0pt, parsep=0pt]

\item 31 years of Doppler monitoring of 51 Peg from the OHP/ELODIE, Lick/Hamilton, Keck/HIRES, and APF/Levy spectrographs reveal signatures of an additional distant companion in the system from a shallow radial acceleration.  The best-fit solution includes curvature with $\dot{\gamma}$ and $\ddot{\gamma}$ values of --1.47$^{+0.01}_{-0.01}$ m $\text{s}^{-1}$ yr$^{-1}$ and 0.0300$^{+0.0003}_{-0.0003}$ m $\text{s}^{-1}$ yr$^{-2}$.

\item The long-term RV signal is driven by the Hamilton dataset.  An analysis of RV standard stars using Hamilton and HIRES data shows that the linear trend observed in the 16-year Hamilton dataset is comparable in scale to the drifts measured in the RV standard stars, making it difficult to rule out instrumental systematics as the origin of the observed radial acceleration.  On the other hand, Hamilton RVs have been able to accurately recover shallow RV trends such as the $\sim$2 m s$^{-1}$ yr$^{-1}$ acceleration from the distant brown dwarf Gl 758 B, indicating that the similar slope from 51 Peg can plausibly be real.

\item A 25-year baseline of absolute astrometry from Hipparcos and Gaia combined with deep imaging from Keck/NIRC2 and HST/WFPC2 rule out stellar companions out to $\approx$390 AU and massive brown dwarfs out to $\approx$40 AU.

\item Between planet b and the potential B/c companion, the 51 Peg planetary system is quite bare.  After removing these signals from the RVs, we can rule out Jovian-mass planets out to 10 AU with high confidence, and Saturns and sub-Saturns out to $\approx$0.5 AU.

\item Over several decades, 51 Peg has exhibited extremely low chromospheric and coronal activity levels, which is in line with its old age of $\sim$5--9~Gyr.  Based on empirical relations connecting activity cycles and rotation periods of solar analogs, the long-term RV signal is most consistent with being
dynamical in origin, as an activity cycle would imply an implausibly long rotation period ($\sim$219 days) and an age ($\sim$200 Gyr) exceeding a Hubble time.
 
\item The combination of these separate, yet complementary, datasets suggests that a giant planet---51Peg c---or brown dwarf companion---51 Peg B---may be present in the system with a mass of $m_p \sin i$ = 10$^{+5}_{-9}$ $M_\mathrm{Jup}$ orbiting at a distance of $a$ = 37$^{+15}_{-20}$ AU.

\item If the observed radial acceleration is real, the origin of 51~Peg~b may be tied to a distant companion (51~Peg~B/c) through high-eccentricity tidal migration following past dynamical interactions. On the other hand, if the acceleration is an instrumental artifact, the absence of additional companions favors a disk-driven migration scenario for the inner hot Jupiter.  We encourage continued long-term RV and astrometric monitoring as well as high-contrast imaging to distinguish between these scenarios.

\end{itemize}

\section{acknowledgments}
We thank the referee for insightful comments and  suggestions that helped improve the content of this manuscript. We thank Jayne Birby for insightful conversations on the possible orbital configurations of the outer companion. B.P.B. acknowledges support from the National Science Foundation grant AST-1909209, NASA Exoplanet Research Program grant 20-XRP20$\_$2-0119, and the Alfred P. Sloan Foundation.
This research has made use of the NASA Exoplanet Archive, which is operated by the California Institute of Technology, under contract with the National Aeronautics and Space Administration under the Exoplanet Exploration Program and \citet{Ochsenbein2000}, an online database with sources collected by the Centre de Données de Strasbourg (CDS).
E.G. acknowledges support from NASA award 80NSSC22K0295 (TESS Guest Observer Program Cycle 4).

This work was supported by a NASA Keck PI Data Award, administered by the NASA
Exoplanet Science Institute. Data presented herein were obtained at the W. M. Keck
Observatory from telescope time allocated to the National Aeronautics and Space
Administration through the agency’s scientific partnership with the California Institute of
Technology and the University of California. Some of the data presented herein were obtained at Keck Observatory, which is a private 501(c)3 non-profit organization operated as a scientific partnership among the California Institute of Technology, the University of California, and the National Aeronautics and Space Administration. The Observatory was made possible by the generous financial support of the W. M. Keck Foundation. 
The authors wish to recognize and acknowledge the very significant cultural role and reverence that the summit of Maunakea has always had within the Native Hawaiian community. We are most fortunate to have the opportunity to conduct observations from this mountain. 
We thank Ken and Gloria Levy, who supported the construction of the Levy Spectrometer on the Automated Planet Finder. We thank the University of California and Google for supporting Lick Observatory and the UCO staff for their dedicated work scheduling and operating the telescopes of Lick Observatory. Support for this work was provided by NASA through the NASA Hubble Fellowship grant HST-HF2-51574 awarded by the Space Telescope Science Institute, which is operated by the Association of Universities for Research in Astronomy, Inc., for NASA, under contract NAS5-26555.

This research has made use of the VizieR catalog access tool,
CDS, Strasbourg, France (doi:10.26093/cds/vizier). The
original description of the VizieR service was published in
\citet{Ochsenbein2000}. This work has made use of data from the European Space Agency (ESA) mission {\it Gaia} (\url{https://www.cosmos.esa.int/gaia}), processed by the {\it Gaia} Data Processing and Analysis Consortium (DPAC,
\url{https://www.cosmos.esa.int/web/gaia/dpac/consortium}). Funding for the DPAC has been provided by national institutions, in particular the institutions participating in the {\it Gaia} Multilateral Agreement.

\facility{Automated Planet Finder (Levy), Keck:I (HIRES), Keck:II (NIRC2), HST (WFPC2)}

\software{\texttt{numpy} \citep{Harris2020}, \texttt{matplotlib}  \citep{Hunter2007}, \texttt{corner} \citep{Foreman-Mackey2016}, \texttt{emcee} \citep{Foreman-Mackey2013}, \texttt{ArviZ} \citep{arviz2019}, \texttt{pandas} \citep{reback2020pandas}; \texttt{ethraid} \citep{VanZandt2024};  \texttt{radvel} \citep{Fulton2018}; \texttt{species} \citep{Stolker2020}; and \texttt{RVSearch} \citep{Rosenthal2021}}

\clearpage
\appendix
\section{Joint Posterior Distributions }\label{sec:Additional_Statistical_Tests} 

\begin{figure}[h!]
\begin{center}
{\includegraphics[width=\textwidth,height=0.7\textheight,keepaspectratio]{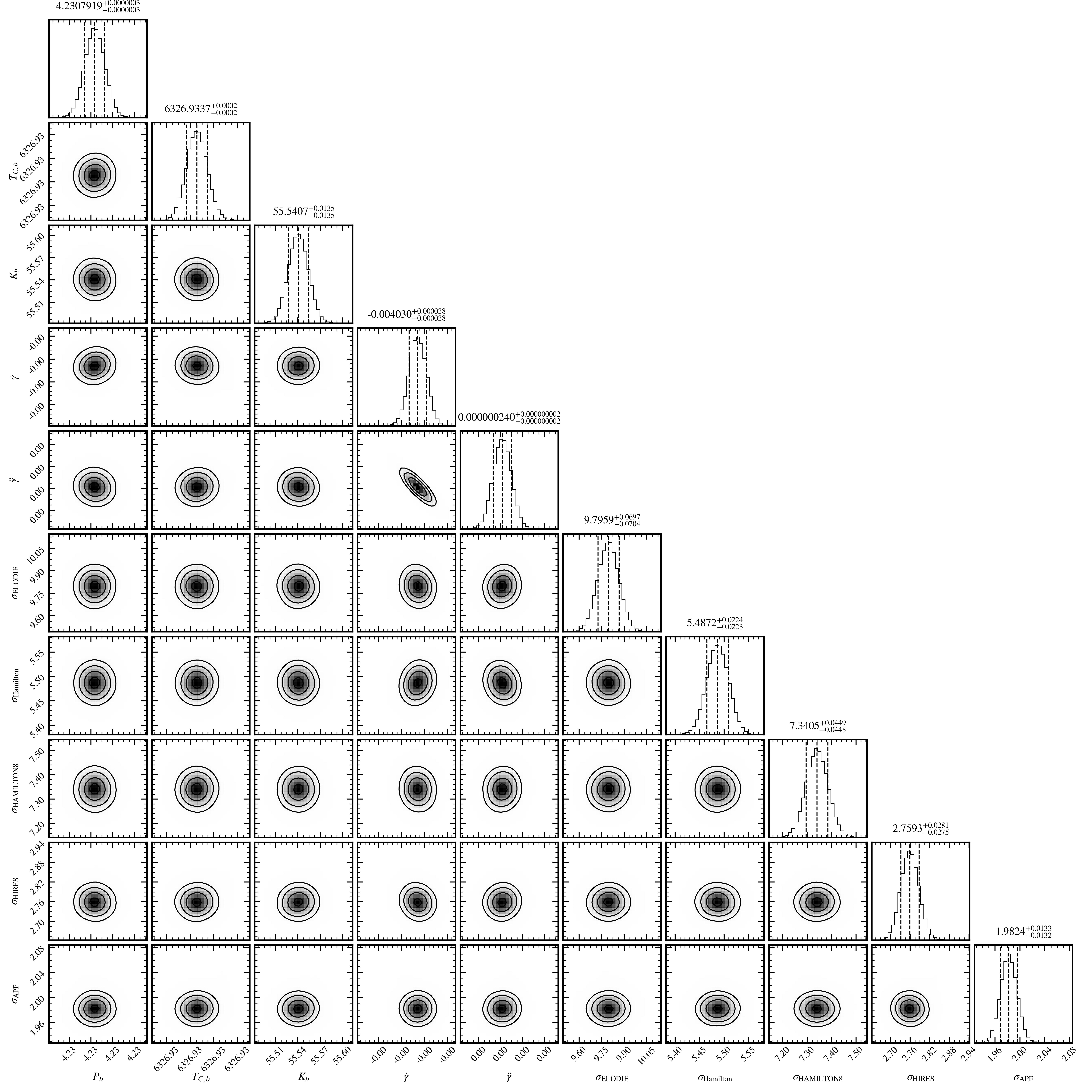}}
\caption{Joint posterior distributions and covariance between model parameters for the Keplerian fit of 51 Peg b and the quadratic trend from 51 Peg B/c. The orbital period ($P_b$, in days), time of inferior conjunction ($T_{C,b}$, in BJD$-2450000$),  semi-amplitude ($K_b$, in m s$^{-1}$), linear acceleration coefficient ($\dot{\gamma}$, in m $\text{s}^{-1}$ d$^{-1}$), quadratic acceleration coefficient ($\ddot{\gamma}$, in m $\text{s}^{-1}$ d$^{-2}$), and jitter terms ($\sigma_\mathrm{ELODIE}$, $\sigma_\mathrm{Hamilton}$, $\sigma_\mathrm{Hamilton8}$, $\sigma_\mathrm{HIRES}$, and $\sigma_\mathrm{APF}$, in m s$^{-1}$) are displayed.} 
\label{fig:corner_plots}
\end{center}
\end{figure}

Here we show the joint posterior distributions of model parameters for the Keplerian-plus-long-term quadradic trend---the statistically favored model that we tested. Figure \ref{fig:corner_plots} illustrates the correlation between all pairs of fitted parameters as a corner plot.  There is little covariance among model parameters except for $\dot{\gamma}$ and $\ddot{\gamma}$, which exhibit a strong negative relationship.  The corresponding best-fit RV solution is displayed in Figure \ref{fig:combined_rv_plot}.

\newpage 

\section{Alternative Models of the RV Data $\&$ BIC Analysis}\label{sec:Additional_RV_Observations}  
\begin{figure}[!htb]
\begin{center}
{\includegraphics[width=0.92\linewidth]{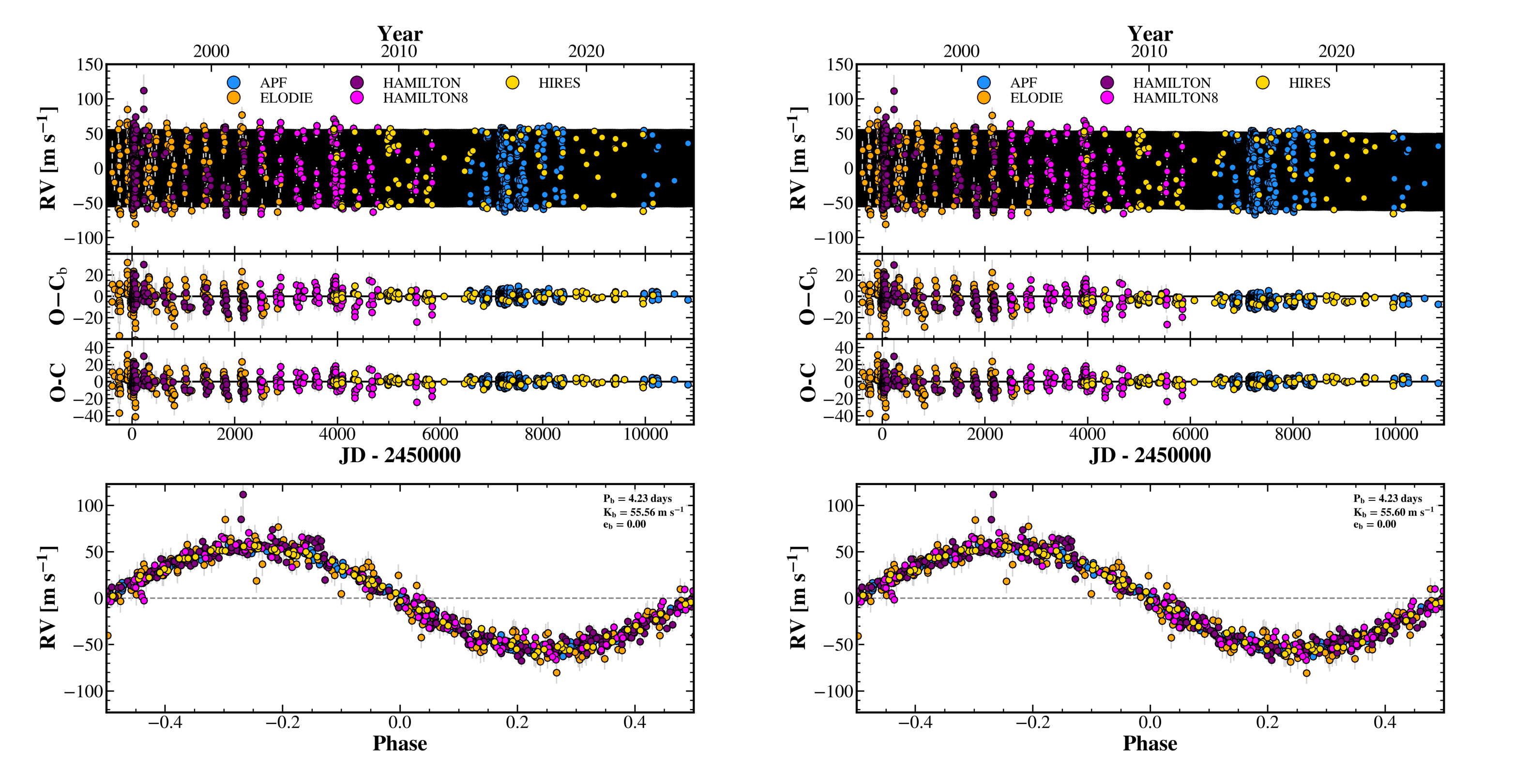}}
\caption{Left: The single-planet Keplerian orbit (51 Peg b) and no radial accelerations in the model. Right: The single-planet Keplerian orbit (51 Peg b) and a first order linear acceleration. Neither model is statistically preferred over the curvature fit shown in Figure \ref{fig:combined_rv_plot}. } 
\label{fig:Additional_RV_Fits}
\end{center}
\end{figure}

Figure \ref{fig:Additional_RV_Fits} shows the additional model fits to the RV data that are not statistically favored. The Keplerian-only solution yields a BIC value of 5653.65, and there are clear trends in the residuals. The Keplerian-plus-linear trend is a better fit, with a BIC of 5653.45, but structure is also present in the residuals. The favored model with a BIC of 5627.85 includes quadratic curvature and is shown in Figure~\ref{fig:combined_rv_plot}.

\newpage

\section{Search for Periodic Signals in the RV Residuals}\label{sec:periodogram_search} 
\begin{figure}[!htb]
\begin{center}
{\includegraphics[height=0.27\textheight,keepaspectratio]{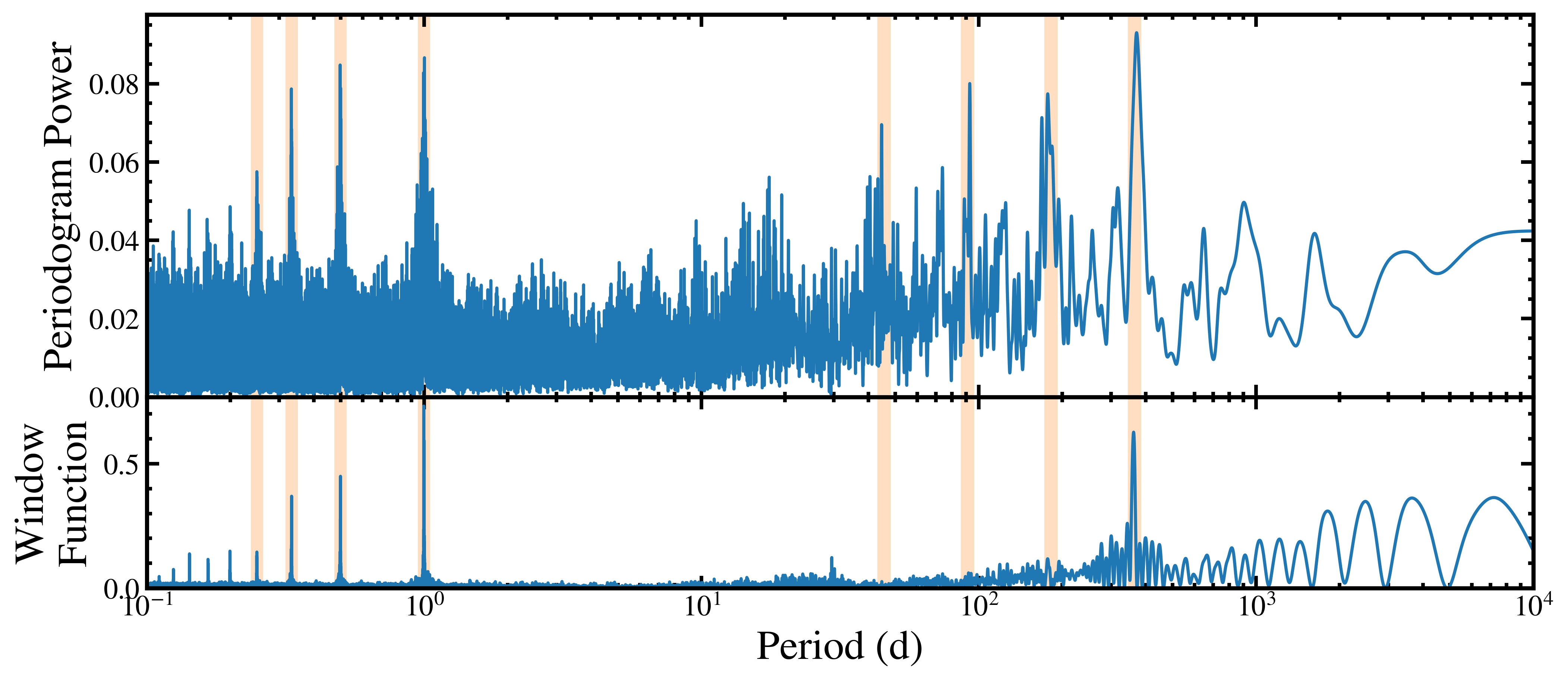}}
\caption{Top: Lomb–Scargle periodogram of the 51 Peg RV data after removing the signal of the inner hot Jupiter, 51 Peg b, as well as the long-term quadratic trend. Bottom: Spectral window
function over the period range. The orange vertical bars correspond to daily or annual aliases of the Earth's orbit at periods of 1 day, day/2, day/4, day/8, 1 year, year/2, year/4, and year/8. The strong peaks visible in the periodogram of the RV data and the spectral window function match the aliases of the Earth's orbit and rotation. There are no additional strong periodic signals found.} 
\label{fig:periodogram_search}
\end{center}
\end{figure}

Figure \ref{fig:periodogram_search} displays a Generalized Lomb-Scargle periodogram \citep{Zechmeister2009} over the frequency range $0.0001{-}10.0~\mathrm{d^{-1}}$ ($0.1{-}10000$ days) to search for any additional periodicity in the residuals of the RV timeseries data. Many strong peaks are present, but all coincide with daily or annual aliasing patterns. 

\newpage

\section{Analysis of Hamilton RV Standards}\label{sec:RV_Standard_Analysis}  
\begin{figure}[!htb]
\begin{center}
{\includegraphics[height=0.7\textheight,keepaspectratio]{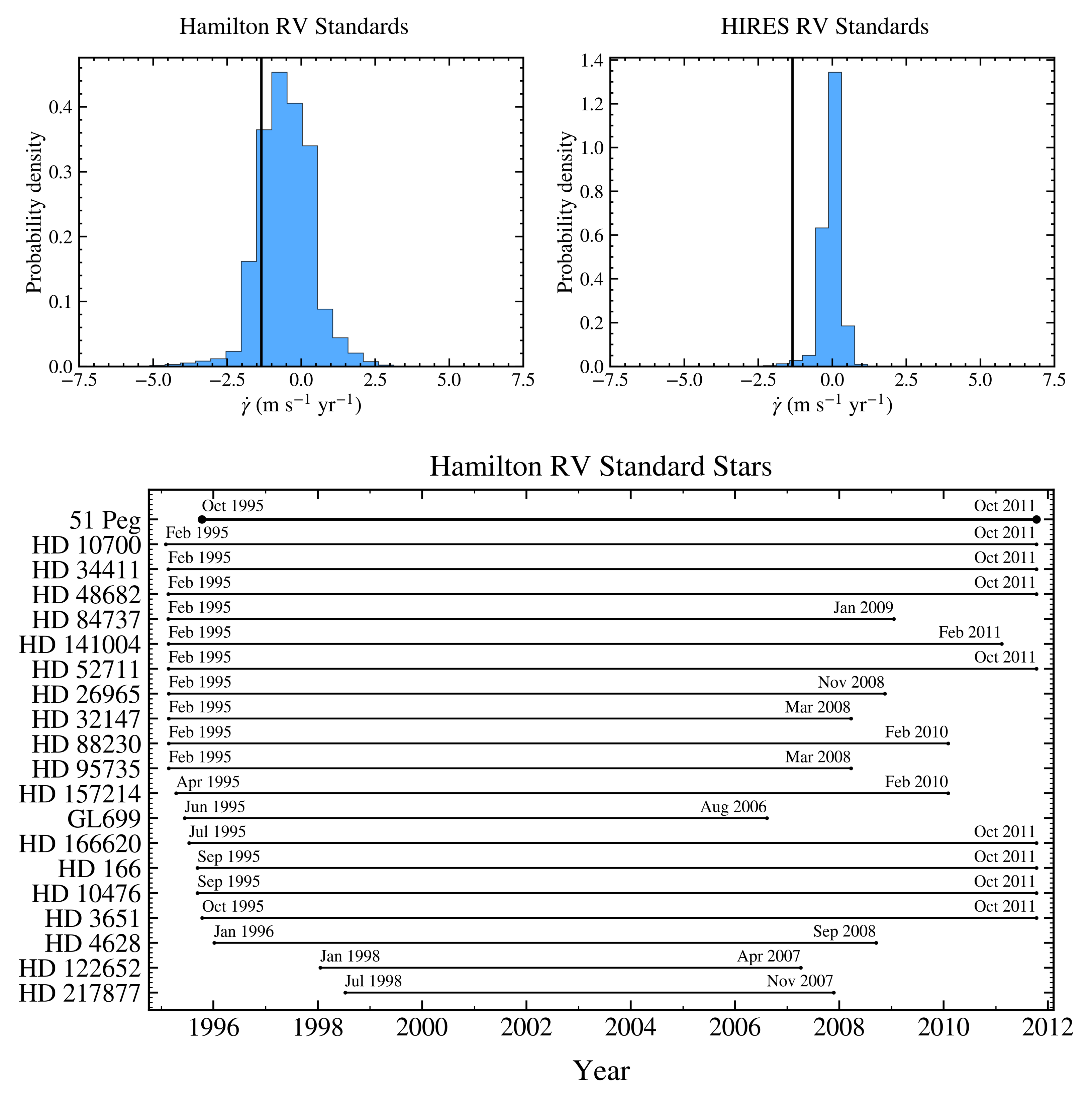}}
\caption{Top Left: The distribution of drifts drawn from Monte Carlo realizations of the Hamilton data and the associated errors assuming normal distributions for each of the 19 stars in our analysis. The solid black line represents the MAP value of the RV trend in the Hamilton dataset. Top right: Distribution of slopes from linear fits of the same RV standard stars with the HIRES data.  Bottom: Baselines of the Hamilton RV standard Doppler measurements used in this analysis compared to the baseline of Hamilton RVs for 51 Peg.} 
\label{fig:Hamilton_RV_Standards}
\end{center}
\end{figure}

Here, we compare the behavior of 19 RV standard stars observed with both the Hamilton and HIRES spectrographs (\citealt{Fischer2014}; \citealt{Rosenthal2021}).  For each dataset, we fit a linear relation to the RVs to assess long-term stability of the stars themselves, and more broadly to establish whether the Hamilton RVs shows trends compared to the more precise and stable HIRES RVs.  The distribution of measured RV drifts of these stable stars, along with their observational baselines relative to 51 Peg, are presented in Figure~\ref{fig:Hamilton_RV_Standards}. In Table~\ref{tab:rv_standards}, we list each RV standard star along with the corresponding RV drift measured in the Hamilton and HIRES datasets.

For the Hamilton datasets, we measure a mean and 1$\sigma$ uncertainty of $\mu_\mathrm{Hamilton} = -0.53$ m s$^{-1}$ yr$^{-1}$ and $\sigma_\mathrm{Hamilton} = 0.90$ m s$^{-1}$ yr$^{-1}$, respectively.  For the HIRES datasets, the mean and 1$\sigma$ uncertainty are $\mu_\mathrm{HIRES} = -0.05$ m s$^{-1}$ yr$^{-1}$ and $\sigma_\mathrm{HIRES} = 0.32$ m s$^{-1}$ yr$^{-1}$, respectively.  The HIRES observations confirm that the RV standard are flat and stable over timescales of at least 1--2 decades.  The Hamilton RVs for the same stars show an average negative drift and spread of $\mu_\mathrm{HIRES}$ - $\mu_\mathrm{Hamilton}$ = 0.48 $\pm$ 0.96 m s$^{-1}$ yr$^{-1}$.  The broadening of the HIRES distribution function amounts to an inflation of $\sqrt{\sigma_\mathrm{Hamilton}^2 - \sigma_\mathrm{HIRES}^2}$ = 0.84 m s$^{-1}$ yr$^{-1}$.  We conclude that long-terms trends from the Hamilton RVs between about --2 m s$^{-1}$ yr$^{-1}$ and +2 m s$^{-1}$ yr$^{-1}$ are consistent with RV stable stars and should be treated with caution.

\begin{deluxetable}{ccccccccc}[!htb]
\renewcommand\arraystretch{0.7}
\tabletypesize{\small}
\setlength{\tabcolsep}{0.1cm}
\tablewidth{0pt}
\tablecaption{RV Standards\label{tab:rv_standards}}
\tablehead{
\colhead{Name} &
\colhead{$N_\mathrm{Hamilton}$} &
\colhead{$\Delta$$t$} &
\colhead{$\dot{\gamma}_{\mathrm{Hamilton}}$} &
\colhead{$\bar{\sigma}_{\dot{\gamma}}$} &
\colhead{$N_\mathrm{HIRES}$} &
\colhead{$\Delta$$t$} &
\colhead{$\dot{\gamma}_{\mathrm{HIRES}}$} &
\colhead{$\bar{\sigma}_{\dot{\gamma}}$} \\
\colhead{} &
\colhead{} &
\colhead{(years)} &
\colhead{(m s$^{-1}$ yr$^{-1}$)} &
\colhead{(m s$^{-1}$ yr$^{-1}$)} &
\colhead{} &
\colhead{(years)} &
\colhead{(m s$^{-1}$ yr$^{-1}$)} &
\colhead{(m s$^{-1}$ yr$^{-1}$)}
}
\startdata
GJ 699 & 29 & 11.16 & 0.53 &  0.69 & 228 & 21.34 & 0.39 & 0.04 \\
HD 10476 & 48 & 16.09 & -0.43 &  0.26 & 421 & 23.41 & -0.28 & 0.05 \\
HD 10700 & 537 & 16.69 & 0.39 &  0.09 & 1071 & 19.41 & -0.02 & 0.02 \\
HD 122652 & 27 & 9.21 & -1.77 &  1.42 & 26 & 20.30 & 0.20 & 0.38 \\
HD 141004 & 235 & 15.98 & -1.44 &  0.14 & 374 & 22.23 & 0.02 & 0.05 \\
HD 157214 & 125 & 14.80 & -0.69 &  0.23 & 91 & 22.44 & 0.01 & 0.06 \\
HD 166 & 21 & 16.09 & 0.88 &  0.84 & 45 & 12.93 & -0.78 & 0.62 \\
HD 166620 & 248 & 16.24 & -0.81 &  0.23 & 143 & 22.77 & -0.09 & 0.06 \\
HD 217877 & 58 & 9.37 & -0.15 &  0.82 & 31 & 17.02 & -0.17 & 0.12 \\
HD 26965 & 78 & 13.73 & -0.91 &  0.25 & 598 & 18.19 & 0.19 & 0.04 \\
HD 32147 & 24 & 13.08 & 0.00 &  0.29 & 572 & 15.34 & 0.10 & 0.02 \\
HD 34411 & 259 & 16.65 & -1.30 &  0.15 & 208 & 19.22 & 0.06 & 0.06 \\
HD 3651 & 138 & 15.99 & -0.62 &  0.27 & 188 & 23.31 & -0.07 & 0.14 \\
HD 4628 & 36 & 12.69 & -0.13 &  0.26 & 290 & 13.17 & 0.11 & 0.06 \\
HD 48682 & 74 & 16.65 & -1.57 &  0.32 & 132 & 13.23 & -0.17 & 0.35 \\
HD 52711 & 42 & 16.64 & -0.56 &  0.28 & 265 & 22.59 & -0.27 & 0.06 \\
HD 84737 & 65 & 13.91 & -1.51 &  0.22 & 108 & 13.23 & 0.06 & 0.11 \\
HD 88230 & 40 & 14.95 & 0.25 &  0.26 & 41 & 11.22 & 0.04 & 0.24 \\
HD 95735 & 42 & 13.08 & -0.24 &  0.36 & 320 & 22.77 & -0.19 & 0.03 \\
\enddata
\end{deluxetable}

\newpage

\section{Analysis of the $S$-value Activity Indictor}\label{sec:RV_Standard_Analysis}  
\begin{figure}[!htb]
\begin{center}
{\includegraphics[width=\linewidth]{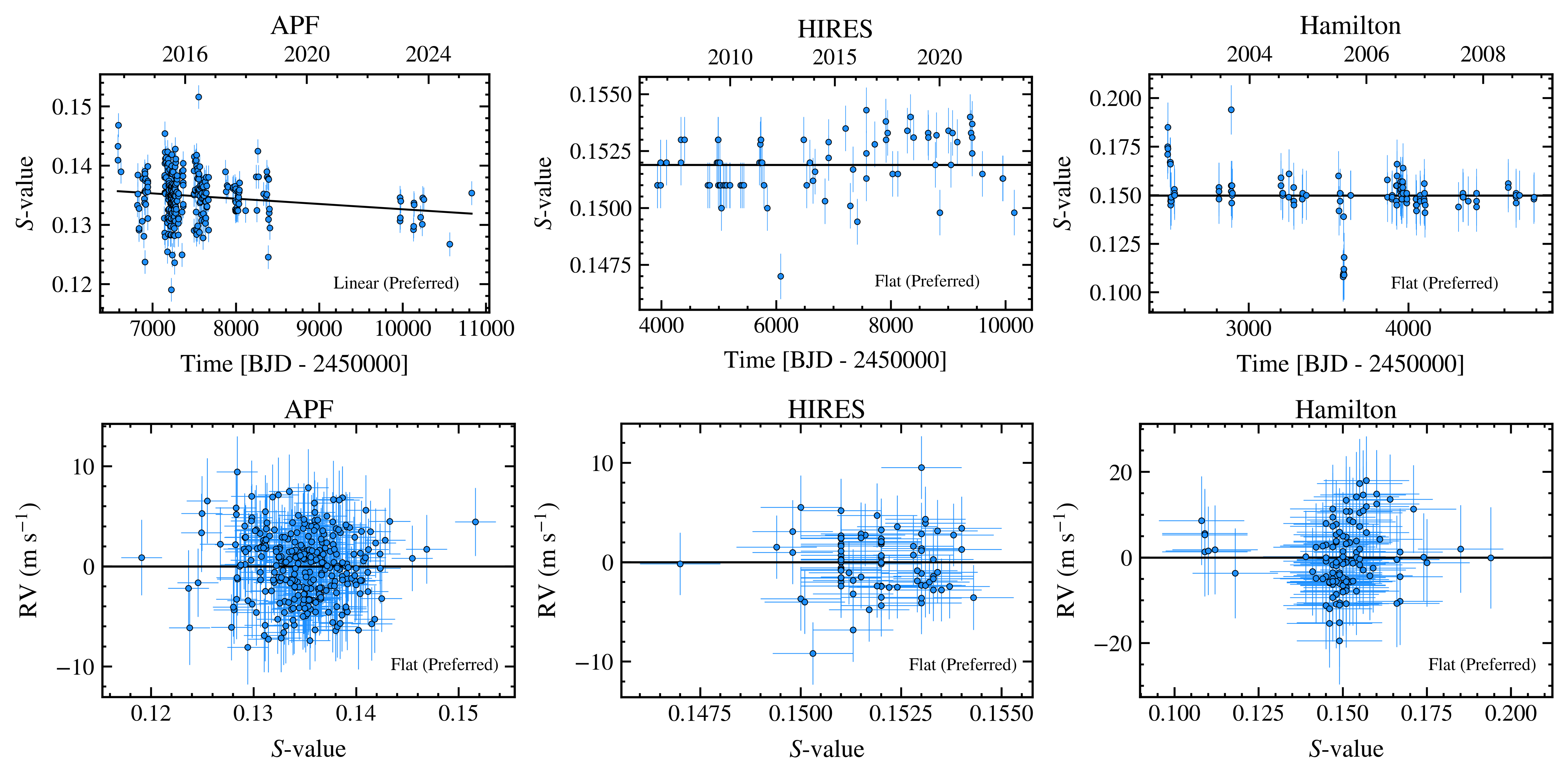}}
\caption{Top: $S$ values as a function of time.  Potential long-term linear trends in the activity indicator measurements are assessed for all 3 datasets. The best-fit model is displayed in the bottom-right corner of each panel; the APF points to a trend, while the HIRES and Hamilton datasets are most consistent with a flat (unchanging) values over time. Bottom: A comparison between the $S$ values and RV measurements.  No significant correlation is evident for any of the three datasets.} 
\label{fig:S_value_time_series}
\end{center}
\end{figure}

We analyze the $S$-value activity measurements from APF, HIRES, and a subset of the Hamilton time series in Figure~\ref{fig:S_value_time_series}. For each dataset, we fit a linear trend in time and compare it to a constant (flat) model using BIC values. A linear trend is statistically preferred in the APF dataset ($\Delta$BIC = 23.1) with a slope of $\dot{{S}}$ =  --0.000326 $\pm$ 0.000121 yr$^{-1}$, while the HIRES ($\Delta$BIC = 6.44) and Hamilton ($\Delta$BIC = 1.71) datasets favor a flat model.  Although the APF $S$-index measurements seem to change over time, this is not reflected in the RVs, for which a flat model is preferred.

We also compute correlation coefficients between the $S$-values and  RVs and similarly compare linear and flat models using BIC values for each model. The correlation coefficients are $-0.03$, $0.01$, and $0.04$ for the APF, HIRES, and Hamilton datasets, respectively, with corresponding $p$-values of $0.56$, $0.90$, and $0.68$. These results indicate no statistically significant correlations are present, and we therefore fail to reject the null hypothesis that the RVs and activity indicators are independent. A $\Delta$BIC comparison between linear and constant models indicates that the constant (flat) model is preferred for all three datasets, with $\Delta$BIC values of 5.97, 4.41, and 4.68 for APF, HIRES, and Hamilton, respectively.

\newpage

\section{Contrast Curves}\label{sec:contrast_curves} 
We report 5$\sigma$ contrast curves in contrast ($\Delta$mag) or sensitivity (apparent magnitude) as a function of angular separation from imaging observations. Table~\ref{tab:Lprime_VVC_Contrast} lists $L'$-band contrasts from the Keck/NIRC2 Vector Vortex Coronagraph observations, Table~\ref{tab:Jc_Contrast} lists $J_\mathrm{cont}$ contrasts from NIRC2, Table~\ref{tab:H_Contrast} lists $H$-band contrasts from NIRC2, and Table~\ref{tab:WFPC2_Contrast} lists the HST/WFPC2 F814W sensitivity limits.

\begin{deluxetable}{cccc}[htb!]
\renewcommand\arraystretch{0.7}
\tabletypesize{\small}
\setlength{\tabcolsep}{0.1cm}
\tablewidth{0pt}
\tablecaption{Keck/NIRC2 $L'$-band Contrast Curve.\label{tab:Lprime_VVC_Contrast}}
\tablehead{\colhead{Separation} & \colhead{$\Delta\mathrm{mag}$} & \colhead{$a$} & \colhead{Mass} \\ 
\colhead{(\arcsec)} & \colhead{(mag)} & \colhead{(AU)} & \colhead{($M_\mathrm{Jup}$)}
}
\startdata
0.19 & 4.87 &  2.90 & 183.92 \\
0.22 & 4.78 &  3.49 & 190.42 \\
0.26 & 4.85 &  4.08 & 185.04 \\
0.30 & 5.68 &  4.66 & 128.36 \\
0.34 & 6.74 &  5.25 & 88.76 \\
0.38 & 7.21 &  5.84 & 81.75 \\
0.41 & 7.68 &  6.43 & 75.00 \\
0.45 & 8.14 &  7.02 & 77.07 \\
0.49 & 8.40 &  7.61 & 76.28 \\
0.53 & 8.90 &  8.20 & 74.77 \\
\enddata
\tablecomments{This table is published in its entirety in machine-readable format.}
\end{deluxetable}

\begin{deluxetable}{cccc}[htb!]
\renewcommand\arraystretch{0.7}
\tabletypesize{\small}
\setlength{\tabcolsep}{0.1cm}
\tablewidth{0pt}
\tablecaption{Keck/NIRC2 $J_\mathrm{cont}$-band contrast curve.\label{tab:Jc_Contrast}}
\tablehead{\colhead{Separation} & \colhead{$\Delta\mathrm{mag}$} & \colhead{$a$} & \colhead{Mass} \\ 
\colhead{(\arcsec)} & \colhead{(mag)} & \colhead{(AU)} & \colhead{($M_\mathrm{Jup}$)}
}
\startdata
0.07 & 2.55 &  1.15 & 577.16 \\
0.13 & 4.29 &  2.08 & 287.77 \\
0.20 & 5.18 &  3.06 & 192.28 \\
0.26 & 5.79 &  4.05 & 154.10 \\
0.33 & 5.98 &  5.06 & 142.06 \\
0.39 & 6.56 &  6.05 & 105.76 \\
0.46 & 7.11 &  7.06 & 93.26 \\
0.52 & 7.37 &  8.05 & 91.34 \\
0.58 & 7.71 &  9.04 & 88.72 \\
0.65 & 7.74 &  10.05 & 88.48 \\
\enddata
\tablecomments{This table is published in its entirety in machine-readable format.}
 \end{deluxetable}

\begin{deluxetable}{cccc}[htb!]
\renewcommand\arraystretch{0.7}
\tabletypesize{\small}
\setlength{\tabcolsep}{0.1cm}
\tablewidth{0pt}
\tablecaption{Keck/NIRC2 $H$-band contrast curve.\label{tab:H_Contrast}}
\tablehead{\colhead{Separation} & \colhead{$\Delta\mathrm{mag}$} & \colhead{$a$} & \colhead{Mass} \\ 
\colhead{(\arcsec)} & \colhead{(mag)} & \colhead{(AU)} & \colhead{($M_\mathrm{Jup}$)}
}
\startdata
0.40 & 6.77 &  6.20 & 97.82 \\
0.60 & 7.47 &  9.31 & 89.33 \\
0.80 & 7.62 &  12.41 & 88.06 \\
1.00 & 8.25 &  15.51 & 82.47 \\
1.20 & 9.00 &  18.61 & 75.80 \\
1.40 & 9.34 &  21.71 & 74.41 \\
1.60 & 9.77 &  24.82 & 74.48 \\
1.80 & 10.10 &  27.92 & 73.54 \\
2.00 & 10.33 &  31.02 & 72.85 \\
2.20 & 10.70 &  34.12 & 71.69 \\
\enddata
\tablecomments{This table is published in its entirety in machine-readable format.}
\end{deluxetable}

\begin{deluxetable}{cccc}[htb!]
\renewcommand\arraystretch{0.7}
\tabletypesize{\small}
\setlength{\tabcolsep}{0.1cm}
\tablewidth{0pt}
\tablecaption{HST/WFPC2 F814W-band Sensitivity Curve.\label{tab:WFPC2_Contrast}}
\tablehead{\colhead{Separation} & \colhead{Sensitivity\tablenotemark{$a$}} & \colhead{$a$} & \colhead{Mass} \\ 
\colhead{(\arcsec)} & \colhead{(mag)} & \colhead{(AU)} & \colhead{($M_\mathrm{Jup}$)}
}
\startdata
1.09 & 11.12 &  16.94 & 83.27 \\
1.23 & 10.64 &  19.05 & 85.78 \\
1.36 & 11.81 &  21.17 & 79.70 \\
1.50 & 11.98 &  23.29 & 78.19 \\
1.64 & 12.58 &  25.41 & 76.64 \\
1.77 & 12.70 &  27.52 & 76.34 \\
1.91 & 13.08 &  29.64 & 75.37 \\
2.05 & 13.12 &  31.76 & 75.25 \\
2.18 & 13.22 &  33.87 & 75.00 \\
2.32 & 13.34 &  35.99 & 74.68 \\
\enddata
\tablenotetext{a}{Apparent F814W magnitude (Vega system).}
\tablecomments{This table is published in its entirety in machine-readable format.}
\end{deluxetable}

\newpage

\bibliography{sample63}{}
\bibliographystyle{aasjournal}

\end{document}